\documentclass[
  pre, 
  preprint,
  floats,    
  amsmath,
]{revtex4} 
\usepackage[dvips]{graphicx}
\begin{document}
 \title{Mesoscopic theory for size- and charge- asymmetric ionic systems.
 I. Case of extreme asymmetry}
 \author { A.Ciach$^{1}$, W.T. G\'o\'zd\'z$^{1}$  and G.Stell$^2$}
\affiliation{$^{1}$Institute of Physical Chemistry, Polish Academy of
 Sciences
 01-224 Warszawa, Poland \break
$^2$Department of Chemistry, State University of New York
 Stony Brook, NY 11 794-3400, USA}
\date{\today} 
\begin{abstract}
A mesoscopic theory for the primitive model of ionic systems is developed
for arbitrary size, $\lambda=\sigma_+/\sigma_-$, and charge,
$Z=e_+/|e_-|$, asymmetry.  Our theory is an extension of the theory we
developed earlier for the restricted primitive model. The case of
extreme asymmetries $\lambda\to\infty$ and $Z \to\infty$ is studied in some
detail in  a mean-field approximation. The phase diagram and correlation
functions are obtained in the asymptotic regime $\lambda\to\infty$ and $Z
\to\infty$, and for infinite dilution of the larger ions (volume
fraction $n_p\sim 1/Z$ or less). We find a coexistence between a very
dilute 'gas' phase and a crystalline phase in which the macroions
form a bcc structure with the lattice constant $\approx
3.6\sigma_+$. Such coexistence was observed experimentally in
deionized aqueous solutions of highly charged colloidal particles.
\end{abstract}
\maketitle                                                         
\section{Introduction}
For many years, theoretical studies of phase behavior in ionic
solutions have been focused mainly on the special case of the restricted
primitive model (RPM), in which half of  equal-sized charged hard spheres
carry positive charge and half carry negative charge of equal magnitude,
with the ions assumed to be dissolved in a structureless solvent 
\cite{stell:76:0,fisher:94:0,outhwaite:04:0,stell:95:0}.
   Even the simplest real ionic solutions have some degree of size
   asymmetry, but with some notable exceptions 
\cite{onsager:68:0,stell:68:0,stell:95:00} a common tacit
   assumption has been that the effects of weak and moderate asymmetry
   in both size and charge is not important to phase behavior.
   Recently, the size and charge asymmetric case has drawn increasing
   attention
   \cite{romero:00:0,yan:01:0,yan:02:0,panag:02:0,cheong:03:0,rescic:01:0,linse:01:0,kalyuzhnyi:00:0,zuckerman:01:0,artyomov:03:0,aqua:04:0,raineri:00:0}.
   Most extensions beyond the RPM are based either on the
   Debye-H\"uckel theory and Poisson-Boltzmann equation, or on the
   mean spherical approximation.  These theories
   \cite{kalyuzhnyi:00:0,zuckerman:01:0,artyomov:03:0,aqua:04:0,raineri:00:0}
   as well as simulations
\cite{romero:00:0,yan:01:0,yan:02:0,panag:02:0,cheong:03:0,rescic:01:0,linse:01:0},
 are  typically limited to the case of small differences in sizes and
  charges. Only in very recent simulations have  moderate
\cite{cheong:03:0} and large \cite{rescic:01:0,linse:01:0}
asymmetries been studied. Moreover, these theories are all "classical" [i.e.,
mean-field-like], and none of them are designed to describe the
special Ising-like behavior that is known to characterize the
primitive model in its critical region.  The development of a theory
that does describe that behavior was sketched by one of us in
 Ref.\onlinecite{stell:99:0a}
and further developed in  Ref.\onlinecite{stell:92:0} and  Ref.\onlinecite{stell:95:0a}.
 A field-theoretic method that
also yields the correct Ising-like behavior was given by Ciach and
Stell in Ref.
\onlinecite{ciach:00:0} and further developed in our subsequent 
papers \cite{ciach:02:0,ciach:05:0,ciach:04:0,ciach:04:1}.
The mesoscopic theory described in this paper is an extension  of the
theory given there, and reduces to it when applied to the restricted
primitive model.

In the case of extreme asymmetry (charge- and diameter ratios between
the two kinds of ions tend to infinity) the PM potentials describe
highly charged colloidal particles suspended in a structureless
solvent containing one kind of counterions and no coions. 
The physical
properties of such a system are significantly different than those of the usual
electrolytes. Highly asymmetric systems exhibit an interesting phase
behavior which is neither fully described nor understood, but it is
clearly quite different than that of  the RPM. In particular, formation of a
colloidal bcc crystal with large inter-particle separation coexisting
with voids
\cite{ise:83:0,ise:99:0,arora:98:0}, various crystals formed by oppositely 
charged colloidal particles \cite{blaaderen:05:0}, and other anomalies
\cite{arora:98:0,tata:97:0,ito:94:0} have been observed. 
The experimental findings suggest the existence of effective attractions
between like-charged macroions as a possible explanation of the
observed phase behavior. The classic  Derjaguin-Landau-Verwey-Overbeek (DLVO)
 theory
\cite{derjaguin:41:0,verwey:43:0}, however, predicts purely 
repulsive interactions between the like-charged colloids.  In
recent approaches  geometrical effects, as well as
fluctuations and correlations  are included
\cite{warren:00:0,barbosa:04:0,belloni:97:0,groot:91:0,messina:00:0,allahyarov:98:0,barbosa:04:0,lowen:92:0,yu:04:0}.
The approaches are based e.g. on integral equations
\cite{belloni:97:0,groot:91:0}, density functionals 
\cite{lowen:92:0,barbosa:04:0,yu:04:0}, and variational methods 
\cite{warren:00:0}. The effective attractions appear
in a  modified DLVO theory \cite{sogami:84:0}, and will also
result from the  explicit inclusion of various effects, such as 'charge
 regulation'
\cite{belloni:97:0}, excluded volume ('Coulomb depletion')
\cite{allahyarov:98:0}, metastable states \cite{messina:00:0}  etc.
 On the other hand, 'volume terms' considered in Ref. \onlinecite{roij:99:0}
 lead to a phase separation for purely repulsive interactions. 
 Some of the mentioned works are questioned by authors of the other
 papers. It goes beyond the scope of our paper to discuss the above
 approaches in more detail; extensive lists of recent works and
 discussions can be found in
 Ref.\onlinecite{warren:00:0,tamashiro:03:0,barbosa:04:0,yu:04:0}. Despite
 impressive progress, the experimentally observed void - bcc crystal
 coexistence has not been predicted, and the issue is still
 controversial. In theoretical approaches to colloidal systems one
 typically assumes extreme size asymmetry between the macro- and
 microions, and the methods differ from those developed for the RPM or
 for the PM with a small asymmetry.

In principle it should be possible to analyze the evolution of phase
diagrams when the size- and charge ratios increase from unity to
infinity. To achieve this goal one needs a theory applicable to
arbitrary size- and charge asymmetries for the PM potentials.  Within
the context of Ornstein-Zernike formalism, one can go quite far in
obtaining the general structure of such a theory, from which a number
of important general results follow, such as the relation between the
charge-charge and density-density correlation lengths, which shows
that they must diverge together in the asymmetric
case~\cite{stell:95:00}.  However, quantitative results for the
thermodynamics and structure of systems of asymmetric ions are very
sensitive to approximations and assumptions
\cite{kalyuzhnyi:00:0,zuckerman:01:0,artyomov:03:0,warren:00:0,barbosa:04:0}.
 In fact one often needs to know the results to make proper
 assumptions, i.e. to identify the physical effects that have to be
 explicitly taken into account (association \cite{kalyuzhnyi:00:0},
 'border zone'\cite{zuckerman:01:0} and cluster \cite{artyomov:03:0}
 formation, 'charge regulation' \cite{belloni:97:0}, 'Coulomb
 depletion'\cite{allahyarov:98:0}, 'volume terms' \cite{roij:99:0}
 etc.).  Recently developed field theory for asymmetric ions, based on
 the Hubbard-Stratonovich transform
 \cite{netz:99:0,netz:00:0,caillol:04:0}, is elegant and in principle
 exact. In practice, however, the phase eqilibria and correlation
 functions can be obtained by using different approximate methods. In
 Ref.\onlinecite{netz:99:0} the size of ions is taken into account through
 the single cutoff in the Fourier space. Since in ionic systems the
 dominant fluctuations are short-range charge-density waves
 \cite{ciach:01:0}, the short-distance properties of the system are
 important, and this approximation may lead to inaccurate
 results. Moreover, the effect of size asymmetry cannot be studied in
 the theory with a single microscopic length. In
 Ref. \onlinecite{caillol:04:0} the hard spheres are taken into account more
 directly, and the charges are smeared inside the spheres to regulate
 the Coulomb potential. Formal expressions and relations are derived
 for arbitrary asymmetry, but in the general case they are very
 complex (and depend on the smearing function) and the author
 restricts the analysis of the thermodynamics and structure to the
 case of equally sized ions.  Reliable theory allowing for a
 determination, with a reasonable effort, of phase equilibria and
 structure in the case of arbitrary asymmetry between the ions has not
 been developed yet.  Therefore the crossover between the case of full
 symmetry and the case of extreme asymmetry is an essentially
 unexplored problem. The PM in the crossover region might be an
 appropriate model for ionic liquids, and it certainly deserves
 attention.

 In a tractable theory simplifying assumptions and approximations
are unavoidable. The key issue is to identify the degrees of freedom
relevant for phase transitions and critical phenomena (i.e. along the
spinodal lines), and to develop a theory which takes them into account
correctly, with the irrelevant degrees of freedom treated in an
approximate way.  In order to describe phase transitions where
ordering occurs at the length scales large compared to molecular
sizes, a coarse-graining procedure, leading to the
Landau-Ginzgurg-Wilson (LGW) approach, has been introduced. The basic
assumption of the LGW theory is that for macroscopic phase separation
the short-wavelength fluctuations, and hence the precise form of
correlations at distances $ r\approx\sigma$ are irrelavant.  In the
case of simple fluids a correlation function can be thought of as
being the sum of two pieces-- the piece that is on the scale of the
distance between particles plus the piece that is on the scale of the
correlation length, which is arbitrarily large close to the spinodal,
and it is only the latter piece that determines universality class and
critical exponents. One can neglect the short-range behavior of
correlation as long as one is in a critical region, [but only then].

In the coarse-grained description one considers deviations from random
distributions of molecules, and it is important to include the
dominant, most probable fluctuations. In simple fluids these correspond
to macroscopic separation, i.e. to
fluctuations with the wavenumber $k\to 0$ in Fourier representation. Because like-charge ions
repell, and oppositely-charge ions attract each other, in ionic
systems charge-ordered
clusters, where positive- and negative-charge ions 
are the nearest-neighbors, are observed in real space~\cite{cheong:03:0,blum:02:0}.
 In Fourier representation the dominant fluctuations are 
charge-density waves
\cite{ciach:00:0,ciach:01:0}. Thus, the fluctuations associated with 
charge ordering in periodic structures should be included in the
coarse-grained description.

The idea of coarse-graining was sucessfuly extended by
Brazovskii~\cite{brazovskii:75:0} and others to soft-matter systems
(liquid crystals, microemulsions, diblock copolymers), where
microphase separation occurs, i.e. periodic phases with a mesoscopic
period of density oscillations may become stable. A mesoscopic period
means a period of order of several molecular diameters or larger. In
this case one expects that the phase equilibria should be
qualitatively correctly described, provided that the fluctuations on
the length scale corresponding to the ordering are included. Again,
 the correlation function consists of a
short-distance piece and of the piece that oscillates on the
mesoscopic scale and decays on the scale of the correlation length,
which is arbitrarily large close to the spinodal. It is the latter
piece that determines the phase transitions, as in simple fluids. The
separation into the short- and long-distance pieces of the correlation
function can be conveniently done by a pole analysis in the complex
Fourier space~\cite{evans:94:0,leote:94:0,stell:99:0,ciach:03:0}. The
dominant pole (or a pair of complex conjugate poles) with the smallest
imaginary part determines the asymptotic large-distance behavior. It
turns out that this dominant pole (or a pair of poles) decribes quite
correctly the correlation function down to the second maximum for
short-range
\cite{evans:94:0} and for Coulombic interactions 
\cite{leote:94:0,ciach:03:0,shim:05:0}. 
In the coarse-grained mesoscopic theories the remainig poles of the
correlation functions are neglected. Note, however that down to the
second maximum in the correlation functions the neglected poles lead
to a small correction to the correlation function, and the results of
the Landau-type, mesoscopic theories work well for such distances,
although for distances $ r\le\sigma$ they are meaningless. To
conclude, if one is interested in the vicinity of the spinodal line
and in the large-distance part of the correlation functions [but only
then], and one wants to take into acount the possiblity of ordering at
distances corresponding to the second maximum of the correlation
function or larger, one can consider the mesoscopic Landau-Brazovskii
theory.

Here we propose to extend the mesoscopic field theory introduced for
the RPM in Ref.\onlinecite{ciach:00:0} to the case of arbitrary size- and
charge asymmetry.  The results of our field theory agree with
simulations
\cite{panag:99:0,diehl:03:0,diehl:05:0,bresme:00:0,vega:03:0} in
continuum-space RPM \cite{ciach:04:0,ciach:02:0}, on the sc and the
fcc lattices \cite{ciach:03:0,ciach:04:0}, and on finely discretized
versions of the former
\cite{ciach:04:0,ciach:04:1}, and also in the presence of additional
short-range  attractive \cite{ciach:01:1} and
repulsive
\cite{ciach:03:0,ciach:04:0} interactions. 
Moreover, the electrostatic free energy has a correct behavior for low
densities, and the exact result in the Debye-H\"uckel limit is
correctly reproduced \cite{ciach:05:0}. So far no example of
qualitatively wrong predictions of the mesoscopic theory has been
found, although in some cases (including the RPM in continuum space)
the effect of fluctuations has to be properly taken into account
\cite{ciach:03:0,ciach:04:0,ciach:04:1}; this can be done
systematically in the perturbation theory.   Foundations of the
mesoscopic theory do not depend on the symmetry properties between the
ionic species.
 On the basis of the results obtained for different
extensions of the RPM one can hope that an extension to the case of
arbitrary asymmetry will also result in a predictive theory yielding
correct results on a semiquantitative level.

In this work we introduce the general framework of the mesoscopic
theory for the PM (sec.2). Next, in sec.3, we focus on the case of
extreme asymmetry, which turns out to be particularly simple. We find
the phase behavior and compare it with experimental results for highly
charged colloidal particles in salt-free water
\cite{ise:83:0,ito:94:0,arora:98:0}.  The agreement is very good.  We also
derive the correlation functions for extremely asymmetric case and
show their forms for various thermodynamic states. We obtain monotonic
decay of correlations for very dilute system, and results consistent
with electric double layer formation for less dilute systems. Near the
transition to the bcc structure the double layer becomes denser and
thiner. We find a pronounced maximum of the colloid correlation
function at distances that agree with experimentally observed ordering
\cite{arora:98:0}. At such distances the clouds of counterions around
the particles do not overlap.  Our results indicate that the theory
developed for arbitrary asymmetry leads to qualitatively correct
predictions in two opposite limiting cases -- fully symmetric (RPM)
\cite{ciach:00:0,ciach:02:0,ciach:03:0,ciach:04:0,ciach:05:0} and extremely asymmetric.
 Hence, we
can expect qualitatively correct results in the crossover region as
well. The results in the case of arbitrary asymmetry will be described
elsewhere.
\section{Mesoscopic theory for the PM}
\subsection{Coarse-graining procedure}
We consider the PM electrolytes with the
diameter and charge ratio between the large and small ions
\begin{equation}
\sigma_+/\sigma_-=\lambda\hskip1cm {\rm and}
\hskip1cm e_+/|e_-|=Z\end{equation}
 respectively (without a loss of generality we assume $e_-=-|e_-|,
 e_+=|e_+|$). For $Z=\lambda=1$ the model reduces to the RPM, and for
 $Z,\lambda\to\infty$ the model describes highly charged colloid
 particles and point-like counterions with a small charge.  In the PM
 the interaction potential of a pair $\alpha,\beta=\pm$ is infinite
 for distances smaller than the sum of radii,
\begin{equation}
\sigma_{\alpha\beta}=(\sigma_{\alpha}+\sigma_{\beta})/2,
\end{equation}
 i.e. we
 assume hard-core repulsions.  The electrostatic potential
 $V_{\alpha\beta}({\bf r}_1-{\bf r}_2)$ between the pair of ions
 $\alpha,\beta $ is
\begin{equation}
\label{Vab}
V_{\alpha\beta}(r)=\frac{e_{\alpha}e_{\beta}}{Dr}
\theta(r-\sigma_{\alpha\beta}),
\end{equation}
where $D$ is the
dielectric constant of the solvent (water).
 The $\theta$-functions above
exclude the contributions to the electrostatic energy coming from
overlapping hard spheres.

In our field-theoretic, coarse-grained approach, we consider local
instantaneous densities of the ionic species, $\rho_{\alpha}({\bf
r})$, i.e. we specify the numbers of ions of both kinds per  mesoscopic
volume $d{\bf r}$. For given densities $\rho_{\alpha}({\bf r})$
precise positions of the ions can be different, and the probability
density $p$ that the local densities assume a particular form
$\rho_{+}({\bf r})$, $\rho_{-}({\bf r})$ is given by
\begin{equation}
\label{pepe}
p[\rho_{\alpha}({\bf r})]=\Xi^{-1}\int_{{\cal S}_p}
 e^{-\beta E({\cal S}_p)} ,
\end{equation}
 where $\beta=(kT)^{-1}$, and where  $T$ and
$k$ are  temperature and the Boltzmann constant
respectively. By $\int_{{\cal S}_p}$ we denote an integral over all
microscopic states ${\cal S}_p$ compatible with the chosen densities
$\rho_{+}({\bf r})$, $\rho_{-}({\bf r})$, and by $ E({\cal S}_p)$ we
denote the energy of the corresponding microstate. The energy of the
microstate ${\cal S}_p$ can be written in the form $ E({\cal S}_p)=
U[\rho_+,\rho_-]+\Delta E_p({\cal S}_p)$, where $U[\rho_+,\rho_-]=
\int_{{\cal S}_p} E({\cal S}_p)/{\cal N}$ is the mean energy
for fixed densities
$\rho_{+}({\bf r})$, $\rho_{-}({\bf r})$,
 and  ${\cal N}= \int_{{\cal S}_p}$ is the number of all microscopic
 states compatible with $\rho_{+}({\bf r})$, $\rho_{-}({\bf r})$. 
We assume that for all
microscopic states compatible with the given local densities the
energy of the whole system is approximately the same, so that
$\beta\Delta E_p({\cal S}_p)\ll 1$.  Hence,
\begin{equation}
\label{ee}
e^{-\beta E({\cal S}_p)}=e^{-\beta
U[\rho_+,\rho_-]}\big[1- \beta\Delta E_p({\cal S}_p) +
\frac{1}{2}(\beta\Delta E_p({\cal S}_p))^2 +...\big] ,
 \end{equation}
and the probability (\ref{pepe}) can be written in the form
\begin{equation}
\label{p}
p=\Xi^{-1}
e^{-\beta U[\rho_+,\rho_-]} \big[{\cal N} + corr\big].
\end{equation}
   The
 correction term is proportional to $\int_{{\cal S}_p}(\beta\Delta
 E_p)^2\ll {\cal N}$ and will be neglected.  Finally, we assume that 
 for particular fields $\rho_{+}({\bf
 r}),\rho_{-}({\bf r})$ the
 electrostatic energy  is given by
\begin{equation}
\label{U}
U[\rho_+,\rho_-]=\frac{1}{2}\int d{\bf r}_1\int d{\bf r}_2  
\rho_{\alpha}({\bf r}_1)V_{\alpha\beta}({\bf r}_1-{\bf r}_2)
\rho_{\beta}({\bf r}_2),
\end{equation}
where the summation convention for Greek indeces is used. 
The fields  $\rho_{\alpha}({\bf r})$
for which $U\to \infty$ occur with the probability  $p\to 0$.
  Hence, in
macroscopic regions the charge neutrality condition
\begin{equation}
\label{neut}
\int d{\bf r}\rho_+({\bf r})e_+=\int d{\bf r}\rho_-({\bf r})|e_-|
\end{equation}
 must be obeyed. One can easily verify that when (\ref{neut}) is
 satisfied for uniform fields $\rho_{\alpha}({\bf r})=const$, then
 $U[\rho_+,\rho_-]=0$. Due to thermal motion the charge neutrality can
 be violated in mesoscopic regions containing a small number of ions.
 The energy (\ref{U}) associated with local deviations from the charge
 neutrality remains finite.

When $\Delta E({\cal S}_p)$
 can be neglected  (i.e. for all microscopic states
compatible with $\rho_{\alpha}({\bf r})$ the energy
 is approximately the same), we can use the Boltzmann formula ${\cal
N}=\exp(\beta TS)$, where by $S$ we denote entropy. 
In an open system the probability is also proportional
to the activities $\exp[\beta(\mu_+N_++\mu_-N_-)]$, where
$N_{\alpha}=\int d{\bf r}\rho_{\alpha}({\bf r}) $ is the number of
ions of the species $\alpha$, and the chemical potentials
$\mu_{\alpha}$ are not independent -- they have to be consistent with
the requirement of the charge neutrality (\ref{neut}).  The above
discussion shows that the local instantaneous densities assume the
form $\rho_{+}({\bf r}),\rho_{-}({\bf r})$ with the probability
density given by
\begin{equation}
\label{Boltz}
p=\Xi^{-1}\exp(-\beta
 \Omega^{MF}[\rho_+,\rho_-]),
\end{equation}
 where $ \Omega^{MF}[\rho_+,\rho_-]$ is the grand potential in the
 system where the local concentrations of the two ionic species are
 constrained to be $\rho_{+}({\bf r})$, $\rho_{-}({\bf r})$. Next we
 assume that the entropy is determined by the hard-core reference
 system with the Helmholtz free energy $F_h=-TS$, and 
 $
\Omega^{MF}[\rho_+,\rho_-]$ is assumed to have the form
\begin{equation}
\label{Omega}
\Omega^{MF}[\rho_+,\rho_-] =F_h[\rho_+,\rho_-] + U[\rho_+,\rho_-] 
-\int d{\bf r} \mu_{\alpha}\rho_{\alpha}({\bf r}).
\end{equation}
For the reference system we assume 
the local density  approximation
  $F_h[\rho_+,\rho_-]=\int d{\bf r} f_h(\rho_+({\bf r}),
\rho_-({\bf r}))$. 
  The $f_h$ consists of the ideal-gas contribution plus the excess
  free-energy density of hard-spheres with different diameters
  $f_h^{ex}$. For example, the Percus-Yevick approximation for
  hard-sphere mixtures \cite{lebowitz:64:0} can be adopted. Because of
  the above assumption, packing effects of hard spheres cannot be
  described in our theory, and in the present form it is not
  applicable to very high densities.  In principle, extensions beyond
  the local density approximation are also possible.

In the field theory introduced above the physical quantities are
obtained by averaging over all fields $\rho_+,\rho_-$ with the
Boltzmann factor (\ref{Boltz}). 
The average densities and the correlation function are
respectively given by
\begin{equation}
\label{roro}
\langle\rho_{\alpha}({\bf r})\rangle= \Xi^{-1}
\int D\rho_+\int D\rho_-
e^{-\beta \Omega^{MF}[\rho_+,\rho_-] }\rho_{\alpha}({\bf r})
\end{equation}
and
\begin{equation}
\label{coco}
G_{\alpha\beta}({\bf r},{\bf r}')=
\langle\rho_{\alpha}({\bf r})\rho_{\beta}({\bf r'})\rangle
-\langle\rho_{\alpha}({\bf r})\rangle\langle\rho_{\beta}({\bf r'})\rangle
\end{equation}
with
\begin{equation}
\label{Gcorr}
\langle\rho_{\alpha}({\bf r})\rho_{\beta}({\bf r'})\rangle=\Xi^{-1}
\int D\rho_+\int D\rho_- 
e^{-\beta \Omega^{MF}[\rho_+,\rho_-] }
\rho_{\alpha}({\bf r})\rho_{\beta}({\bf r}'),
\end{equation}
and
\begin{equation}
\label{Xi}
\Xi=\int D\rho_+\int D\rho_- 
e^{-\beta \Omega^{MF}[\rho_+,\rho_-] }.
\end{equation}
The grand potential $\Omega$ is 
\begin{equation}
\label{grdef}
-\beta\Omega=\log \Xi.
\end{equation}
 In practice we are not able to evaluate the functional integrals in
 Eqs. (\ref{roro}), (\ref{Gcorr}) and (\ref{Xi}), and we need to
 make approximations. In the simplest, mean-field (MF) approximation
 the average values of the local densities are approximated by their
 most probable values, $\rho_{0\alpha}$, and the grand thermodynamic
 potential is approximated by the minimum of
 $\Omega^{MF}[\rho_+,\rho_-]$ at $\rho_{\alpha}=\rho_{0\alpha}$.

 As convenient thermodynamic variables we choose dimensionless number
 density of all ionic species, $s$, and dimensionless temperature
 $T^*=1/\beta^*$, where
\begin{equation}
\label{s}
s=\frac{\pi}{6}(\rho_{0+}^*+\rho_{0-}^*), \hskip0.5cm
\beta^*=\beta\frac{e_+|e_-|}{D\sigma_{+-}},
\end{equation}
and 
\begin{equation}
\rho_{\alpha}^*=\sigma_{+-}^3\rho_{\alpha}.
\end{equation}
Here and below as a length unit we choose
  $\sigma_{+-}$. Because of the charge neutrality,
\begin{equation}
\label{ch_neut}
\rho_{0-}^*=Z\rho_{0+}^*,
\end{equation}
 the volume fraction of ionic species is
\begin{equation}
\label{zeta}
\zeta=\frac{\pi}{6}(\rho_{0+}\sigma_+^3+\rho_{0-}\sigma_-^3)=
\frac{2^3(\lambda^3+Z)}{(1+\lambda)^3(1+Z)}s.
\end{equation}

Let us study the form of $\Omega^{MF}$ in more detail.
For small deviations 
$\Delta\rho^*_{\alpha}({\bf x})=
\rho_{\alpha}^*({\bf x})-\rho_{0\alpha}^*$ 
of the local densities
 from their  most probable
values the grand potential
(\ref{Omega}) can be expanded about its value  $\Omega^{MF}_0$ at 
the minimum,
\begin{equation}
\label{OmF}
\Delta \Omega^{MF}=\Omega^{MF}- \Omega^{MF}_0=\Omega^{MF}_2
+\Omega^{MF}_{int}.
\end{equation}
Here  $\Omega^{MF}_2$  denotes  the Gaussian part of the functional. In
Fourier representation we have
\begin{equation}
\label{OmF2}
\beta\Omega^{MF}_2=\frac{1}{2}\int\frac{d {\bf k}}{(2\pi)^3}
\Delta \tilde\rho^*_{\alpha}(-{\bf k})\tilde C^0_{\alpha\beta}({\bf k})
\Delta \tilde\rho^*_{\beta}({\bf k}) 
\end{equation}
where $\Delta \tilde\rho^*_{\alpha}({\bf k})$ is the Fourier transform
of $\Delta\rho^*_{\alpha}({\bf x})$, and the wave numbers are in
$\sigma_{+-}^{-1}$ units. The second functional derivatives of
 $\Omega^{MF}$,  $\tilde
C^0_{\alpha\beta}({\bf k})$, consist of two terms,
\begin{equation}
\label{Calphabeta}
\tilde C^0_{\alpha\beta}({\bf k})=a_{\alpha\beta} +\beta \tilde
V_{\alpha\beta}(k).
 \end{equation}
The first term is given by
 the corresponding second derivative of $\beta f_h$ 
 taken at $\rho_{\alpha}^*=\rho_{0\alpha}^*$.
 The second term in $\tilde
C^0_{\alpha\beta}({\bf k})$ is the Fourier transform of the potential
(\ref{Vab}),  and we find
\begin{equation}
\label{Vfab}
\beta \tilde V_{\alpha\beta}(k)=
\frac{e_{\alpha}e_{\beta}}{e_+|e_-|}
\frac{4\pi \cos(kr_{\alpha\beta})}{k^2}
\beta^*,
\end{equation}
where $r_{\alpha\beta}=\sigma_{\alpha\beta}/\sigma_{+-}$.
The remaining part of the functional has the expansion
\begin{equation}
\label{omegaint}
\beta\Omega^{MF}_{int}=\int d{\bf r}\Bigg[
\frac{a_{\alpha\beta\gamma}}{3!}\Delta\rho^*_{\alpha}({\bf r})
\Delta\rho^*_{\beta}({\bf r})\Delta\rho^*_{\gamma}({\bf r})+
\frac{a_{\alpha\beta\gamma\delta}}{4!}\Delta\rho^*_{\alpha}({\bf r})
\Delta\rho^*_{\beta}({\bf r})\Delta\rho^*_{\gamma}({\bf r})
\Delta\rho^*_{\delta}({\bf r})\Bigg]+ ...,
\end{equation}
where 
\begin{equation}
\label{aabc}
a_{\alpha\beta\gamma}=
\frac{\partial a_{\alpha\beta}}{\partial\rho^*_{\gamma}}\hskip1cm{\rm and}
\hskip1cm
{a_{\alpha\beta\gamma\delta}}=
\frac{\partial a_{\alpha\beta\gamma}}{\partial\rho^*_{\delta}},
\end{equation}
and the derivatives are taken at $\rho^*_{\alpha}=\rho^*_{0\alpha}$.
For pair-potentials and for the local-density approximation for the
reference system, $\Omega^{MF}_{int}$ is strictly local. The explicit forms
of the coefficients depend on the reference system.
\subsection{Phase transitions}
Let us focus on the boundary of stability of the uniform phase.
The uniform phase is unstable with respect to fluctuations
$\tilde\rho^*_{\alpha}({\bf k})$ when the second functional derivative
of $\Omega^{MF}$ is not positive definite, i.e. $\det \tilde
C^0_{\alpha\beta}(k)<0$. 
The
temperature at the instability with respect to the ${\bf k}$-mode is
thus given by
\begin{equation}
\label{D}
\det \tilde C^0_{\alpha\beta}({\bf k})=0.
\end{equation}
  Boundary of stability with respect to the deviations
  $\Delta\rho^*_{\alpha}\propto\cos({\bf r}\cdot{\bf k})$ from the
  densities $\rho^*_{0\alpha}$ corresponds to ${\bf k}={\bf k}_b$ such
  that the Eq.(\ref{D}) is satisfied first when the temperature is
  decreased. For fixed $\rho^*_{0\alpha}$ the boundary of stability is
  thus given by the maximum of $T^*({\bf k})$ obtained from (\ref{D}),
  therefore ${\bf k}_b$ can be determined from the equations
\begin{equation}
\label{pD}
\frac{\partial (\det \tilde C^0_{\alpha\beta}(k_i))}{\partial k_i}=0.
\end{equation}
Solutions of the set of equations (\ref{D}) and (\ref{pD}) give both,
the wave vector of the critical fluctuations ${\bf k}_b$, and the
spinodal line in the phase space $(s,T^*)$. For temperatures higher
than at the spinodal line the randomly chosen instantaneous local
densities are most probably uniform. For lower temperatures, however,
the randomly chosen instantaneous densities most probably have a form
of planar waves with the wave vector ${\bf k}_b$, or of linear
combinations of such waves with different orientations of the wave
vectors. The amplitudes of the density waves of the ionic species and the 
order of the associated phase transition depend on the
form of $\Omega_{int}$.  In the next section we shall find the phase
transition in the case of extreme asymmetry in MF.
\subsection{Correlation functions}
\label{sec2.2}
Let us consider the structure of the disordered phase,
i.e. the correlation functions for the density deviations from
$\rho^*_{0\alpha}$. In our mesoscopic theory, especially in the local
density approximation, the correlation functions defined in
Eq.(\ref{Gcorr}) are meaningful for distances larger than
$\sigma_{\alpha\beta}$, and in principle we can only expect a
semiquantitative agreement with results of exact theories or
simulations for large distances. The pole analysis of the
correlation functions in Fourier representation shows that in the
mesoscopic theory for the RPM only the dominant poles, characterizing the
long-distance behavior, are present~\cite{ciach:03:1}. On the other
hand, the dominant poles yield quite accurate results down to the second maximum 
of the correlation functions
\cite{leote:94:0,shim:05:0}, and for such distances we can  expect 
semiquantitatively correct results in the colloid limit as well.  However, the functional integrals in
Eq.(\ref{Gcorr}) cannot be calculated exactly.  In practice we are
able to calculate $G_{\alpha\beta}(r)$ in a perturbation expansion in
$\gamma_{2n,m}$.  In the Gaussian approximation, $\Delta\Omega^{MF}=
\Omega^{MF}_2$, i.e. with the term $\Omega^{MF}_{int}$ in
Eq.(\ref{OmF}) neglected, the correlation functions (\ref{Gcorr}) can
be easily calculated by inverting the matrix of second functional
derivatives of $\Omega^{MF}$. In Fourier representation we have thus
\begin{equation}
\label{corinv}
\tilde G^0_{\alpha\beta}({\bf k})
=\big[\tilde {\bf C}^0({\bf k})\big]_{\alpha\beta}^{-1}.
\end{equation}
The functions analogous to pair distribution functions are related to
 $G_{\alpha\beta}$ (Eq.(\ref{Gcorr})) by
\begin{equation}
\label{smallg}
g_{\alpha\beta}(r)=\frac{G_{\alpha\beta}(r)}{\rho^{*}_{0\alpha}
\rho^{*}_{0\beta}}
-\frac{\delta({\bf r})\delta^{Kr}_{\alpha\beta}}{\rho^*_{\alpha}}+1.
\end{equation}
Beyond the Gaussian approximation we expect corrections to the
correlation functions. Their relevance in different thermodynamic
states will be studied in future works. 

\section{Case of extreme asymmetry}
The above model can be applied to a suspension of highly charged
colloid particles in a (salt-free) solvent containing one kind of
point-like counterions.  In the case of extreme asymmetry the
reference system corresponds to a mixture of hard spheres and
point-like species, with the densities constrained according to
Eq.(\ref{ch_neut}).  For $\lambda\to \infty$ the volume fraction
$\zeta$ reduces to the volume fraction $n_p$ of the large species, and
in the asymptotic regime $Z\to\infty$ (such that $Z/\lambda^3\to 0$)
we obtain ((see (\ref{zeta}), (\ref{ch_neut}) and (\ref{s})),
\begin{equation}
\label{zetalim}
\zeta=n_p= 8s/Z.
\end{equation}
 Hence, finite values of the number density, $s=O(Z^0)$, correspond to
 infinite dilution of hard spheres for $Z\to\infty$.   At infinite
 dilution a hard-sphere system can be approximated by an ideal gas. The
 smaller ions are point-like in the limit $\lambda\to \infty$, and
 also behave as an ideal gas. Thus, for $\lambda, Z\to \infty$ we can
 assume that the reference system is just a mixture of ideal-gases.
 For a mixture of ideal-gases
 $a_{\alpha\beta}=\delta^{Kr}_{\alpha\beta}/\rho^*_{0\alpha}$, and
 from Eqs. (\ref{s}) and (\ref{ch_neut}) we obtain
\begin{equation}
\label{a--++}
 a_{--}=\frac{1}{\rho^*_{0-}}=\frac{\pi }{6s} ,
\hskip1cm a_{++}=\frac{1}{\rho^*_{0+}}=\frac{\pi }{6s} Z  ,
\hskip1cm a_{+-}=0.
\end{equation}
Strictly speaking, in our  analysis we assume that
$\lambda\to\infty$ first, and next we consider the asymptotic behavior
of $Z\to\infty$ with $s=O(Z^0)$.  The above asymptotic behavior
 will be referred
to as the colloid limit. From Eq. (\ref{Vfab}) we easily find that in
the colloid limit the electrostatic potentials are
\begin{equation}
\beta\tilde V_{++}(k)=O(Z), \hskip0.5cm 
\beta \tilde V_{+-}(k)=O(Z^0), \hskip0.5cm\beta\tilde V_{--}(k)=O(Z^{-1}).
\end{equation}
 Hence, in the colloid limit we obtain 
\begin{equation}
\label{CCC}
\tilde C_{++}(k)= Z\tilde C_{p}(k),\hskip0.5cm
\tilde C_{+-}(k) =\beta \tilde V_{+-}(k)
=O(Z^0),
\hskip0.5cm\tilde C_{--}(k)=\frac{\pi }{6s}+ O(Z^{-1})
\end{equation}
 and in turn
\begin{equation}
\det \tilde C_{\alpha\beta}(k)= 
 Z\frac{\pi}{6s} \tilde C_{p}(k)+O(1),
\end{equation}
where
\begin{equation}
\label{Cp}
 \tilde C_{p}(k)= \frac{\pi}{6}\Bigg[\frac{1}{s}+
\frac{24\cos(2k)}{k^2}\beta^*\Bigg].
\end{equation}

\subsection{Stability of the disordered phase}
 Instability of the disordered phase, in general given in Eq.(\ref{D}),
 for $Z\gg 1$ is equivalent to 
\begin{equation}
  \big[\tilde C_{p}(k)\frac{\pi }{6s}+O(Z^{-1})\big]=0,
\end{equation}
and for $Z\to\infty$ the latter equation is satisfied when $\tilde
C_{p}= 0$. From Eq.(\ref{CCC}) we obtain that if $\tilde C_{p}=0$,
then $ \tilde C_{++}(k)=0$ for arbitrarily large $Z$.  This means that
in the considered asymptotic regime of $Z\to\infty$ and $n_p=O(Z^{-1})$, the 
fluctuations $
\Delta\tilde \rho_{+}({\bf k})$ can destabilize the uniform phase. 
The line of instability of the
uniform phase with respect to these fluctuations, given by (\ref{D})
and (\ref{pD}), assumes the form
\begin{equation}
 \tilde C_{p}(k)=
0=\partial \tilde C_{p}(k)/\partial k.
\end{equation}
The spinodal line is given by the explicit expression
\begin{equation}
\label{spinmacro}
T^*_b(s)=-\frac{24\cos(2k_b)}{k_b^2}s, \hskip0.5cm 
 \tan(2k_b)=\frac{1}{k_b},
\end{equation}
and we find   $k_b\approx 1.23$ in $\sigma_{+-}^{-1}$ units. 

In order to determine the phase transition associated with the 
spinodal (\ref{spinmacro}), let us consider the asymptotic behavior 
of $\Omega_{int}$  
 (Eq. (\ref{OmF})) for $ Z\to
\infty$
  with $s=O(Z^0)$ and $\lambda\to
\infty$. For the reference system corresponding to
 a mixture of ideal gases the only nonvanishing coefficients in
 Eq.(\ref{OmF}) are (see Eq.(\ref{a--++}))
\begin{equation}
 a_{\alpha \alpha \alpha }=-\frac{1}{\rho^{*2}_{0\alpha}},\hskip1cm
 a_{\alpha \alpha\alpha \alpha }=\frac{2}{\rho^{*3}_{0\alpha}}.
\end{equation}
After using Eqs. (\ref{s}) and (\ref{ch_neut}) we obtain
\begin{eqnarray}
\label{Omasy}
\beta\Delta\Omega^{MF}=\frac{Z}{2}
\int_{\bf k}\Delta\tilde\rho^*_+({\bf k})
\tilde C_p(k)\Delta\tilde\rho^*_+(-{\bf k})+\\ \nonumber
\int_{\bf r}\Big[\frac{Z^2A_3}{3!}\Delta\rho^{*3}_+({\bf r})+
\frac{Z^3A_4}{4!}\Delta\rho^{*4}_+({\bf r})+....\Big]+ O(Z^0),\\ 
\nonumber
\end{eqnarray}
where 
\begin{eqnarray}
\label{AAs}
A_3=-\Bigg(\frac{\pi}{6s}\Bigg)^2,\hskip1cm
 A_4=2\Bigg(\frac{\pi}{6s}\Bigg)^3,
\end{eqnarray}
and by $O(Z^0)$ we denote all remaining contributions to
$\Delta\Omega^{MF}$ with the integrands that remain finite or tend to
zero
 for $Z\to\infty$. All integrands proportional to $\Delta\rho^{*}_-$ 
 turn out to be $O(Z^0)$. For $Z\to\infty$ we
neglect such terms compared to those given in Eq.(\ref{Omasy}). Note
that we again come to the conclusion that the phase transition in the
colloid limit is determined only by the macroion-density fluctuations.
In the second step we rescale the field, $\Delta\tilde\rho^*_+=
\tilde\psi/Z$, and the functional, 
$\Delta\Omega[\Delta\tilde\rho_+,\Delta\tilde\rho_-]
= \Omega_r[\tilde\psi]/Z$, and  we obtain
\begin{eqnarray}
\label{Omren}
\beta\Omega_r[\tilde\psi]=\frac{1}{2}\int_{\bf k}
\tilde\psi({\bf k})\tilde C_p(k)\tilde\psi(-{\bf k})\hskip2cm \\ 
\nonumber
+\frac{A_3}{3!}
\int_{{\bf k}_1}\int_{{\bf k}_2}\int_{{\bf k}_3}\delta(\sum_i^3{\bf k}_i)
\prod_i^3\tilde\psi({\bf k}_i)\\ 
\nonumber
+
\frac{A_4}{4!}
\int_{{\bf k}_1}\int_{{\bf k}_2}\int_{{\bf k}_3}\int_{{\bf k}_4}
\delta(\sum_i^4{\bf k}_i)\prod_i^4\tilde\psi({\bf k}_i),
\nonumber
\end{eqnarray}
where we truncated the expansion in the field at the fourth order
term. Because the cubic term is present, the transition to the phase
with periodic ordering of the particles is first order in MF.
 A similar functional was already studied by Leibler in
the context of block copolymers \cite{leibler:80:0}, and we can
directly use his results. In order to find the stable structure one
considers $\psi$ of a form of linear superpositions of $n$ planar
waves with the wave vectors ${\bf k}^j_b$ having different
orientations (with $j=1,...,n$ and $|{\bf k}^j_b|=k_b$),
\begin{eqnarray}
\label{psssi}
\tilde\psi({\bf k})=\frac{\Phi}{\sqrt n}\sum_j^n
\Big[\delta({\bf k}-{\bf k}_b^j)w_n+\delta({\bf k}+{\bf k}_b^j)w_n^*\Big],
\end{eqnarray}
where $w_n$, $w_n^*$ are complex conjugate and $w_nw_n^*=1$. 
For $\tilde\psi({\bf k})$ of the form (\ref{psssi}) the functional $\Omega_r$ 
(\ref{Omren}) per volume $V$
can be written as
\begin{eqnarray}
\label{Omres}
\beta\Omega_r/V=\tilde C_p(k_b)\Phi^2-\alpha_3\Phi^3+\alpha_4\Phi^4,
\end{eqnarray}
where the geometric factors $\alpha_j$ depend on $n$ and have been
found in Ref. \onlinecite{leibler:80:0} for several structures. For
metastable structures $\Omega_r$ assumes  local minima, 
\begin{eqnarray}
\label{Omder}
\partial\beta\Omega_r/\partial\Phi=0,
\end{eqnarray}
 and the stable structure corresponds to
the global minimum. At the coexistence of the disordered phase with
the periodic structure 
\begin{eqnarray}
\label{Omzero}
\beta\Omega_r=0,
\end{eqnarray}
 and the actual phase transition occurs when the above equation is
 satisfied for this phase which corresponds to the global minimum of
 $\beta\Omega_r$; another words, for the phase which becomes stable,
 Eq.(\ref{Omzero}) is satisfied at the highest temperature. From
 (\ref{Omres}), (\ref{Omder}) and (\ref{Omzero}) we obtain the
 transition line as $\tilde C_p(k_b)=\alpha_3^2/(4\alpha_4)$.  From
 the results of Ref.\onlinecite{leibler:80:0} it follows that in MF the
 disordered fluid coexists with the bcc crystalline structure if
 $A_3\ne 0$. In the bcc arrangement of colloids the wave vectors ${\bf
 k}^j_b$ form edges of a regular tetrahedron, and for our particular
 case $\alpha_3=4A_3/(3\sqrt 6)$ and $\alpha_4=5A_4/8$
 \cite{leibler:80:0}. The above results and Eqs.(\ref{Cp}) and
 (\ref{AAs}) enable us to obtain the explicit expression for the
 transition line in MF
\begin{equation}
\label{bccT}
T^*=-1.063\frac{24\cos(2k_b)}{k_b^2}s\approx 13.1 s\approx 1.64 Zn_p.
\end{equation}
The transition line is given by the above equation  only for  $n_p=O(1/Z)$,
 because our asymptotic analysis here is restricted to $s=O(Z^0)$.  

In the colloid limit it is more convenient to use
$\sigma_+=2\sigma_{+-}$ as a length unit, and $ \sigma_+^{-1}$ as a
wavelength unit. To avoid confusion, the wave numbers in
$\sigma_+^{-1}$ units will be denoted by $ q$. 
In real space the field (\ref{psssi}) 
in the case of the bcc structure is for ${\bf r}=(x,y,z)$
 in suitably chosen coordinate frame given by
\begin{eqnarray}
\label{bbcc}
\Delta\rho_+(x,y,z)\propto\psi(x,y,z)\propto\cos\Big(\frac{q_b(x+y)}{\sqrt 2}\Big)+
\cos\Big(\frac{q_b(x+z)}{\sqrt 2}\Big)+
\cos\Big(\frac{q_b(y+z)}{\sqrt 2}\Big)+ \\ \nonumber
\cos\Big(\frac{q_b(x-y)}{\sqrt 2}\Big)+
\cos\Big(\frac{q_b(x-z)}{\sqrt 2}\Big)+
\cos\Big(\frac{q_b(y-z)}{\sqrt 2}\Big).
 \end{eqnarray}
In Fig.1 the surface $\Delta\rho_+(x,y,z)=0$ is shown.  This surface
separates the regions with the particle density exceeding the average
value from the regions of depleted particle density.  From
(\ref{bbcc}) we see that the lattice constant $a$ of the bcc structure
is related to the critical wavenumber $ q_b$ by $a=2\sqrt 2
\pi/q_b\approx 3.6$ in $\sigma_+$ units.  The distance between
nearest-neighbors in the bcc crystal is $\sqrt 6\pi/q_b\approx
3.12\sigma_+$. 
\begin{figure}
\includegraphics[scale=0.5]{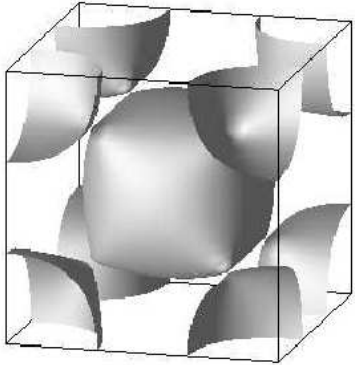}
\caption{The surface $\Delta\rho_+(x,y,z)=0$, separating the regions of
enhanced and depleted density of particles. $\Delta\rho_+(x,y,z)>0$ inside 
the droplets. The cubic unit cell with the 
lattice constant $a=2\sqrt 2\pi/q_b$ is shown.
}
\end{figure}
Let us consider the above transition between the uniform phase and the
bcc crystal for typical (deionized) aqueous systems studied in
Refs.\onlinecite{ise:83:0,arora:98:0,ise:99:0}, where $Z\sim 10^{3} - 10^5$ and
$\lambda \sim 10^3$. From (\ref{s}) we find that room temperature
is $T^*\sim 10^{-1} -10^{-3}$, and   Eq. (\ref{bccT}) gives $n_p\sim
10^{-4}-10^{-8}$ in the uniform phase at the coexistence with the bcc
crystal. Such dilutions  can be recognized as 'voids' in experiments.
 Moreover, in experiments the interparticle distance
 in the bcc crystal was found to be $\approx 3\sigma_+$
\cite{ise:83:0,arora:98:0,ise:99:0}, in  agreement with the results
 of our theory.
 The above behavior is not confirmed by simulations
 \cite{cheong:03:0,rescic:01:0,linse:01:0}, where gas-liquid type
 separation with a critical point, rather than crystallization, has
 been observed. However, in
 Refs.\onlinecite{cheong:03:0,rescic:01:0,linse:01:0} the size- and/or
 charge asymmetry is two-three orders of magnitude smaller than in
 experiments.

The asymptotic theory described above is strictly valid
 only in the colloid limit ($\lambda\to\infty$ first and next the
 asymptotic behavior for $Z\to\infty$ with $n_p \sim 1/Z$ or smaller
 is considered).  Beyond the colloid limit the full set of equations (\ref{D})
 have to be solved. Since there are two
 eigenmodes in the general case, both eigenvalues can vanish, leading
 to two spinodal lines associated with two phase transitions. As will
 be shown in Ref.
\onlinecite{ciach:05:1}, the other spinodal  is associated with a gas-liquid
 separation.  Beyond MF the crystallization may be preempted by the
 gas-liquid separation, as is the case for the
 RPM~\cite{ciach:03:0,ciach:02:0,ciach:04:0}. The latter transition
 corresponds to vanishingly low values of $n_p$ and $T^*$ in the limit
 $Z,\lambda\to\infty$, therefore only the crystallization survives in
 the colloid limit. Thus, although predictions of our theory in the
 colloid limit disagree with the results of simulations (obtained
 beyond that limit) it is plausible that the results of the full
 theory will agree with simulations for appropriate values of the
 asymmetry parameters and for the corresponding regions in the phase
 diagram $(n_p,T^*)$.  It is worth noting that in snapshots shown in
 Ref.\onlinecite{cheong:03:0} clusters separated by 'voids' are
 clearly distinguishable.

\subsection{Gaussian correlation functions}
 For temperatures lower than that given in Eq.(\ref{bccT}) the
 periodic structure is stable. This suggests effective attractions
 between like-charged macroions at the distances $r \sim 3\sigma_+$,
 at least near the transition to the crystalline phase.  In fact
 already the experimental discovery of void-crystal coexistence and
 other anomalies
\cite{arora:98:0}
 inspired a debate on the origin of the effective attraction between
 like charged particles~
\cite{warren:00:0,roij:99:0,arora:98:0,belloni:97:0,allahyarov:98:0,messina:00:0}. 

In the mesoscopic theory instead of effective interactions between the
macroions in the uniform phase we consider the correlation function $
G_{++}({\bf r}_1-{\bf r}_2)$ defined in Eq.(\ref{coco}), and related
to the pair correlation function according to
Eq.(\ref{smallg}). Maxima of $ G_{++}({\bf r}_1-{\bf r}_2)$ indicate
increased probability of finding a pair of colloid particles at the
corresponding positions. In the lowest order, Gaussian approximation
(neglected $\Omega_{int}$ in Eq.(\ref{OmF})) the correlation functions
are given in Eq.(\ref{corinv}), and in the colloid limit (i.e. for
$n_p\sim 1/Z$ and $\lambda,Z\to\infty$) we find
\begin{equation}
\label{corgauss}
\tilde G_{++}(q)
 =\frac{T^*}{4Z} \Bigg[S+\frac{4\pi\cos q}{q^2}\Bigg]^{-1},
\end{equation}
%
\begin{equation}
\label{corgauss--}
\tilde G_{--}(q)=\frac{6s}{\pi}-\frac{4\pi}{Sq^2}
\Bigg(1-\frac{4\pi\sin ^2(q/2)}{Sq^2}\Bigg)\tilde G _{++}(q)
\end{equation}
and
\begin{equation}
\label{corgauss+-}
\tilde G_{+-}(q)=\frac{4\pi\cos (q/2)}{Sq^2}\tilde G _{++}(q),
\end{equation}
where 
\begin{equation}
\label{S}
S=\pi T^*/(24s),
\end{equation}
 and terms $O(Z^{-2})$ have been neglected. Note that the
 $q$-dependent parts of the correlation functions are all of the same
 order $O(Z^{-1})$. The $q$-dependent parts of the functions
 (\ref{corgauss})-(\ref{corgauss+-}) multiplied by $4Z\beta^*$ are
 independent of $Z$, and depend on the thermodynamic state only
 through $S$. In the colloid limit the functions
 $4Z\beta^*G_{\alpha\beta}$ assume universal shapes 
along the straight lines (\ref{S}) in the phase diagram $(s,T^*)$,
 at least in the Gaussian approximation.

From the above
 result and  from Eq.(\ref{ch_neut})  it follows
 that  for $s=O(Z^0)$  the corresponding
$g$-functions (Eq.(\ref{smallg})) are $\tilde g_{++}=O(Z)$,
$\tilde g_{+-}=O(Z^0)$ and $\tilde g_{--}=O(Z^{-1})$.
 Note the strong 
dependence of these functions on the charge asymmetry resulting from the
 difference in the number densities in the charge-neutral system.  

In real-space representation $G_{\alpha\beta}(r)$ can be obtained by
residue method \cite{leote:94:0,ciach:03:1}. All the functions have the same
denominator, hence the same poles determine the decay lengths and
(where applicable) the period of damped oscillations. The form of
$\tilde G_{++}(q)$ is similar to the form of charge-density
correlation function in the RPM, and the latter was studied in
Ref.\onlinecite{ciach:03:1}. From the results of Ref.\onlinecite{ciach:03:1} it
follows that for $S>S_K\approx 11.8$, where $S=S_K$ is known as the
Kirkwood line \cite{leote:94:0}, there are two imaginary poles $ia_1$ and
$ia_2$ in the upper half of the complex plane, and
\begin{equation}
\label{monodecay}
rG_{\alpha\beta}(r)=A^{(1)}_{\alpha\beta}e^{-a_1r}+
A^{(2)}_{\alpha\beta}e^{-a_2r}.
\end{equation}
For $S<S_K$ there are two conjugate complex poles,
$q_1=\alpha_1+i\alpha_0$ and $q_2=-q_1^*$ with $\alpha_0>0$, and
\cite{ciach:03:1}
\begin{equation}
\label{corr++real}
rG_{\alpha\beta}(r)=
{\cal A}_{\alpha\beta} \sin(\alpha_1 r+\theta)e^{-\alpha_0r}.
\end{equation}
The poles can only be found numerically, except near the spinodal line
\cite{ciach:03:1}. Following the analysis of Ref.\onlinecite{ciach:03:1} we
find the characteristic lengths and the amplitudes. The correlation
functions $4Z\beta^*G_{\alpha\beta}$ are shown in Figs. 2-4 for three
different regimes. In Fig.2 we show the correlation function above the
Kirkwood line, i.e. for very dilute systems or for very high
temperatures. Fig.3 corresponds to $S<S_K$, i.e to denser system
and/or lower temperatures, but far from the phase
coexistence. Finally, in Fig.4 we show the correlation functions at
the coexistence with the bcc crystal.
\begin{figure}
\includegraphics[scale=0.3,angle=270]{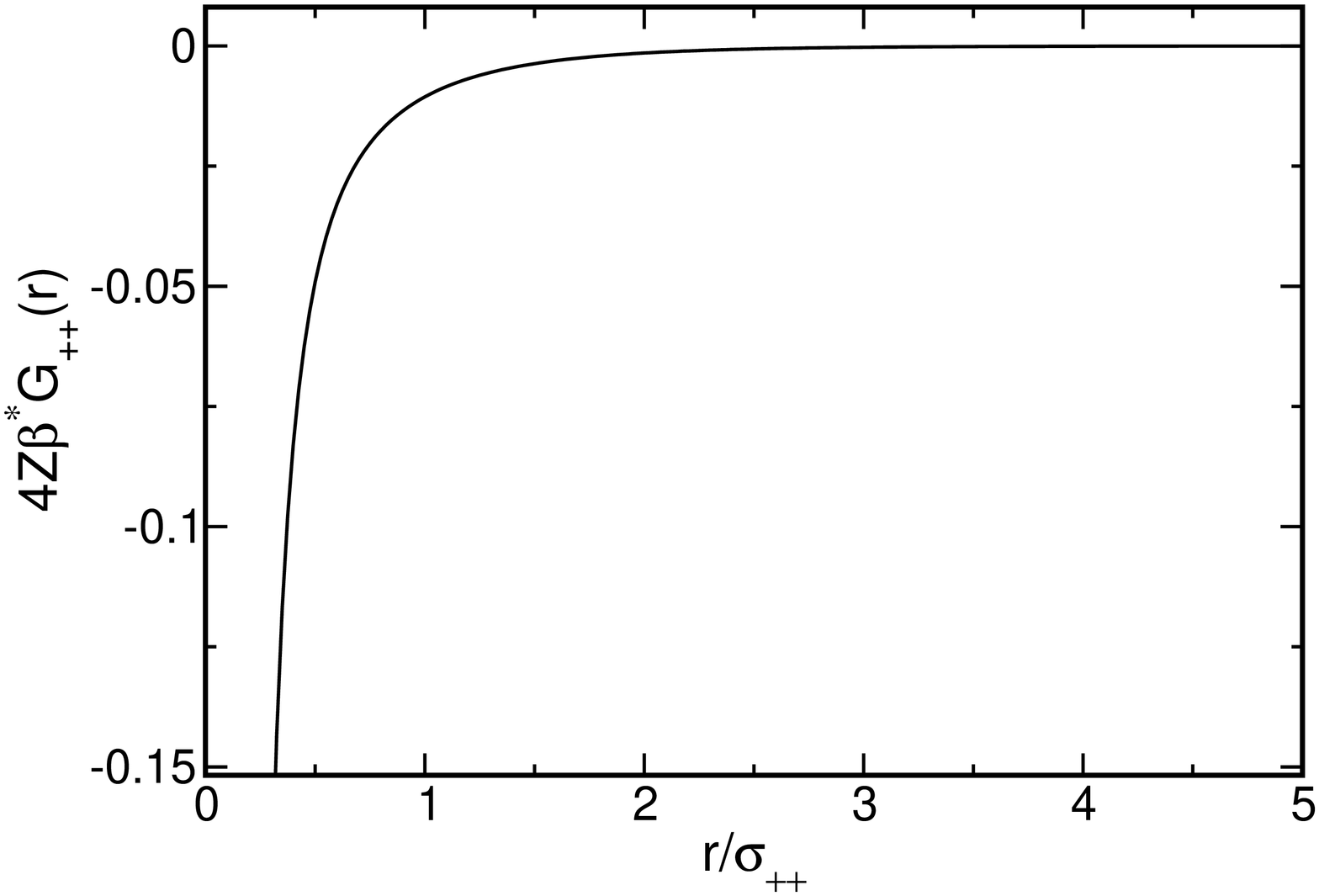}
\includegraphics[scale=0.3,angle=270]{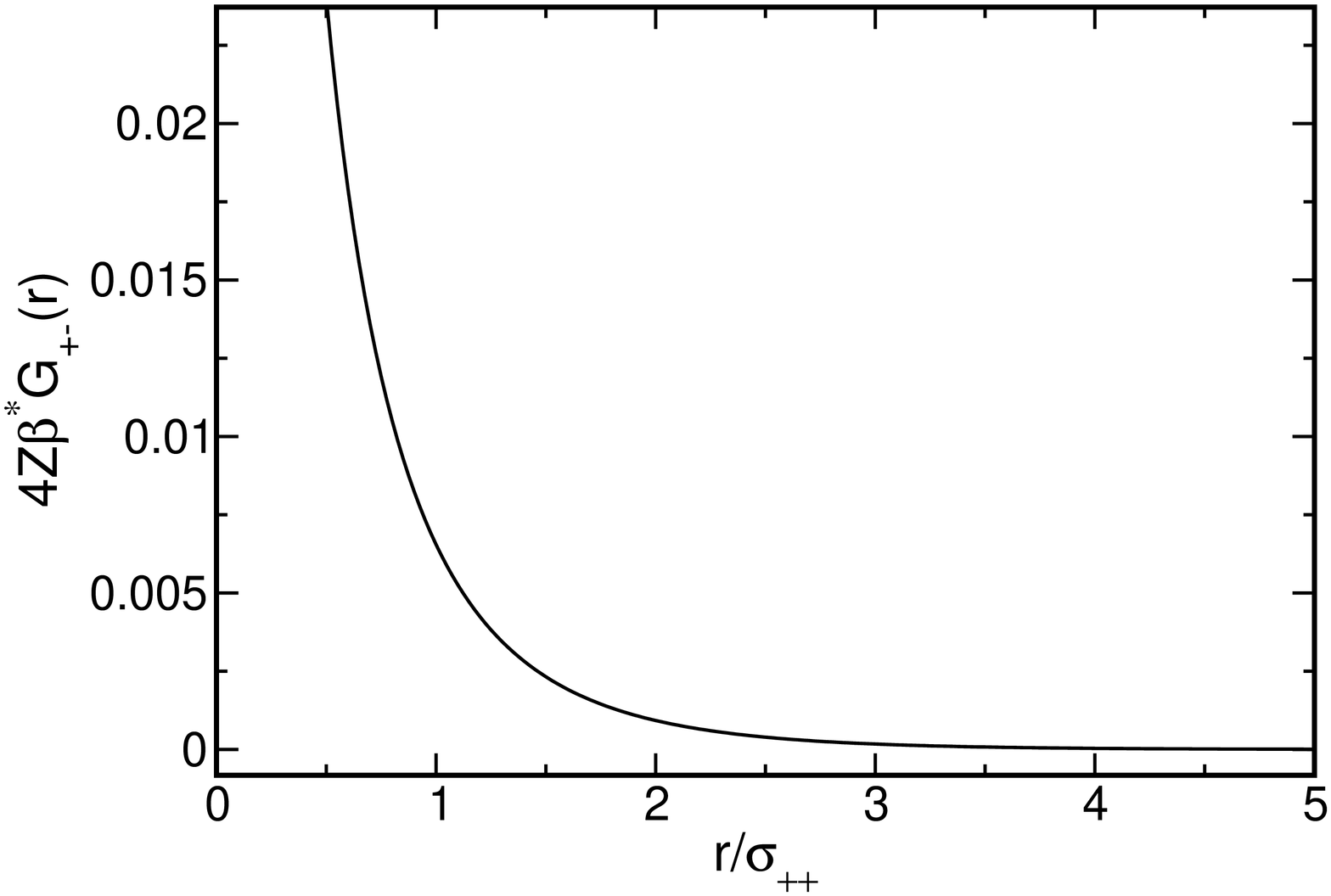}
\includegraphics[scale=0.3,angle=270]{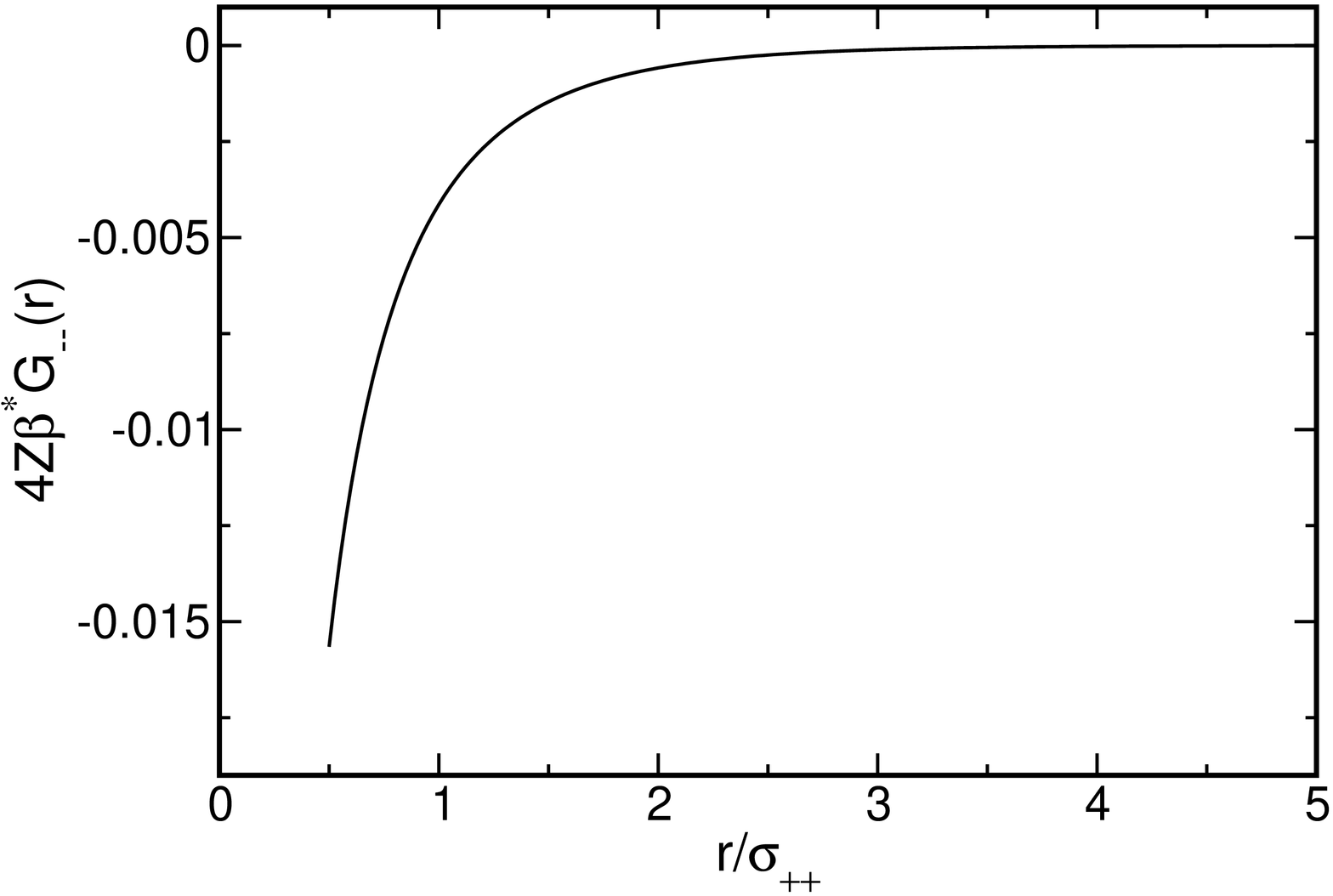}
\caption{Correlation functions $4Z\beta^*G_{\alpha\beta}(r)$ (dimensionless) in
the uniform phase for $S=15$, i.e. for very strong dilutions. The
inverse decay lengths in Eq.(\ref{monodecay}) are $a_1=1.265$ and
$a_2=3.19$. Distance is in units of the particle diameter. As
discussed in sec.~\ref{sec2.2}, results of the mesoscopic theory for
$r\le\sigma$ are not expected to be correct.
}
\end{figure}
\begin{figure}
\includegraphics[scale=0.3,angle=270]{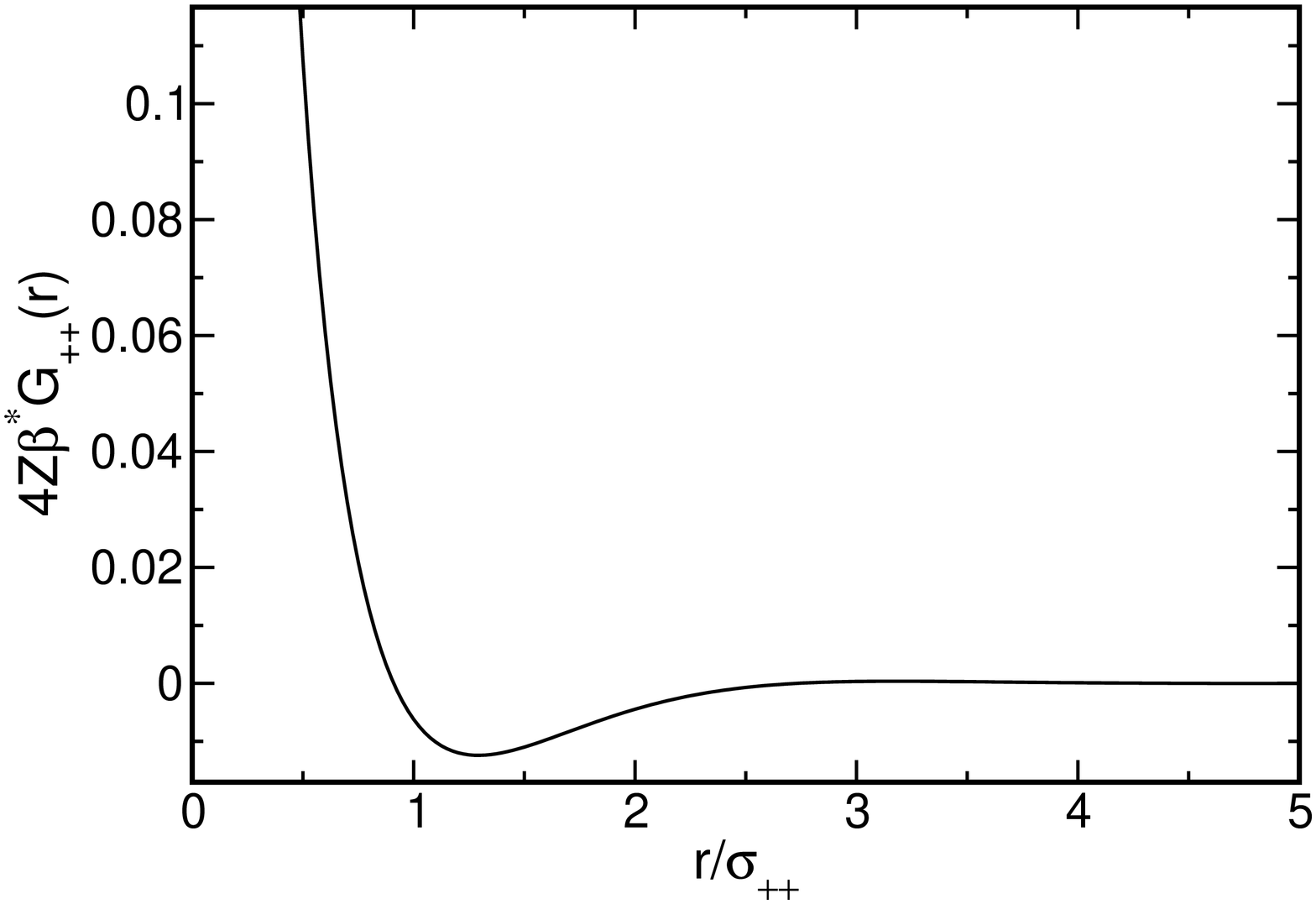}
\includegraphics[scale=0.3,angle=270]{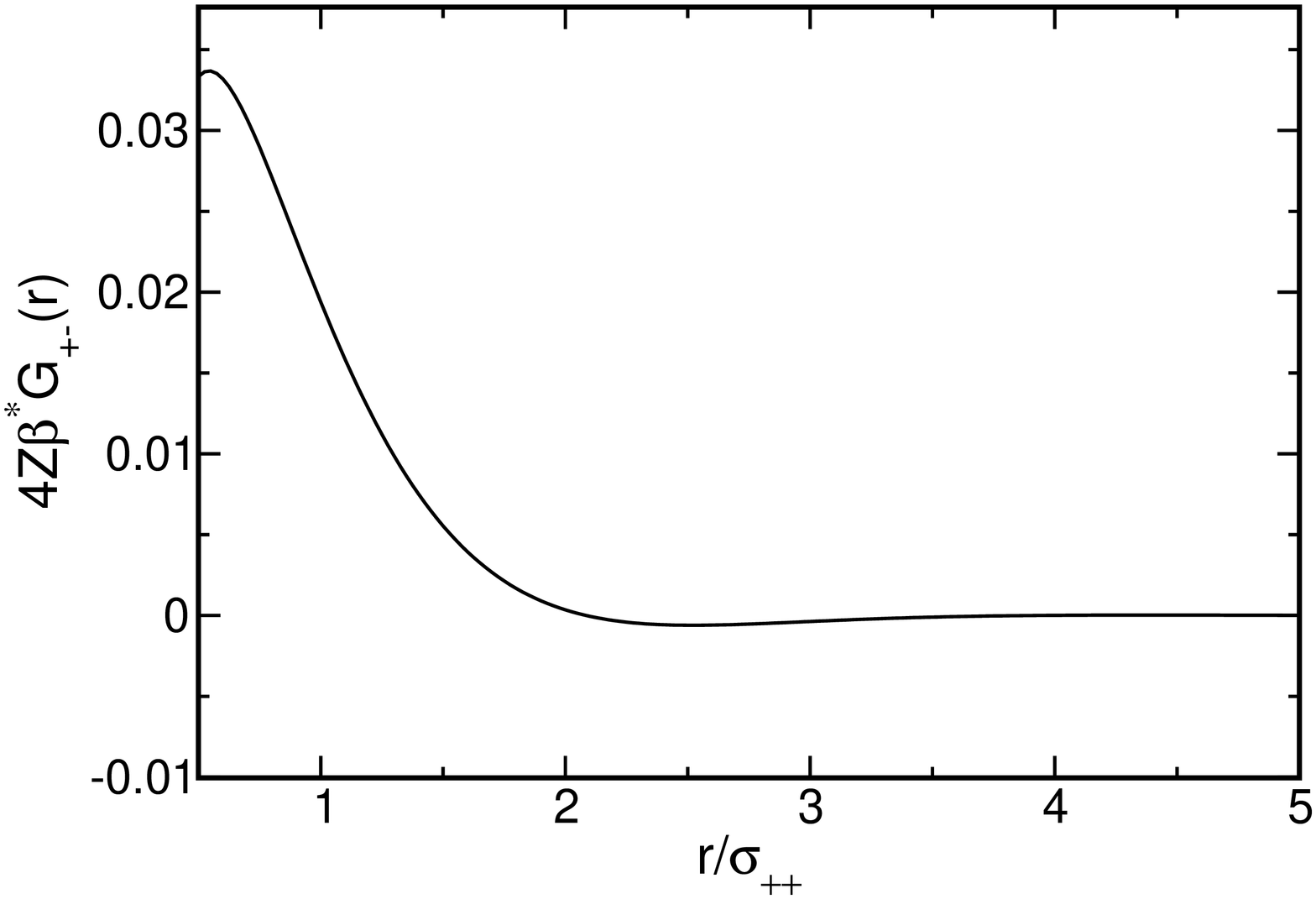}
\includegraphics[scale=0.3,angle=270]{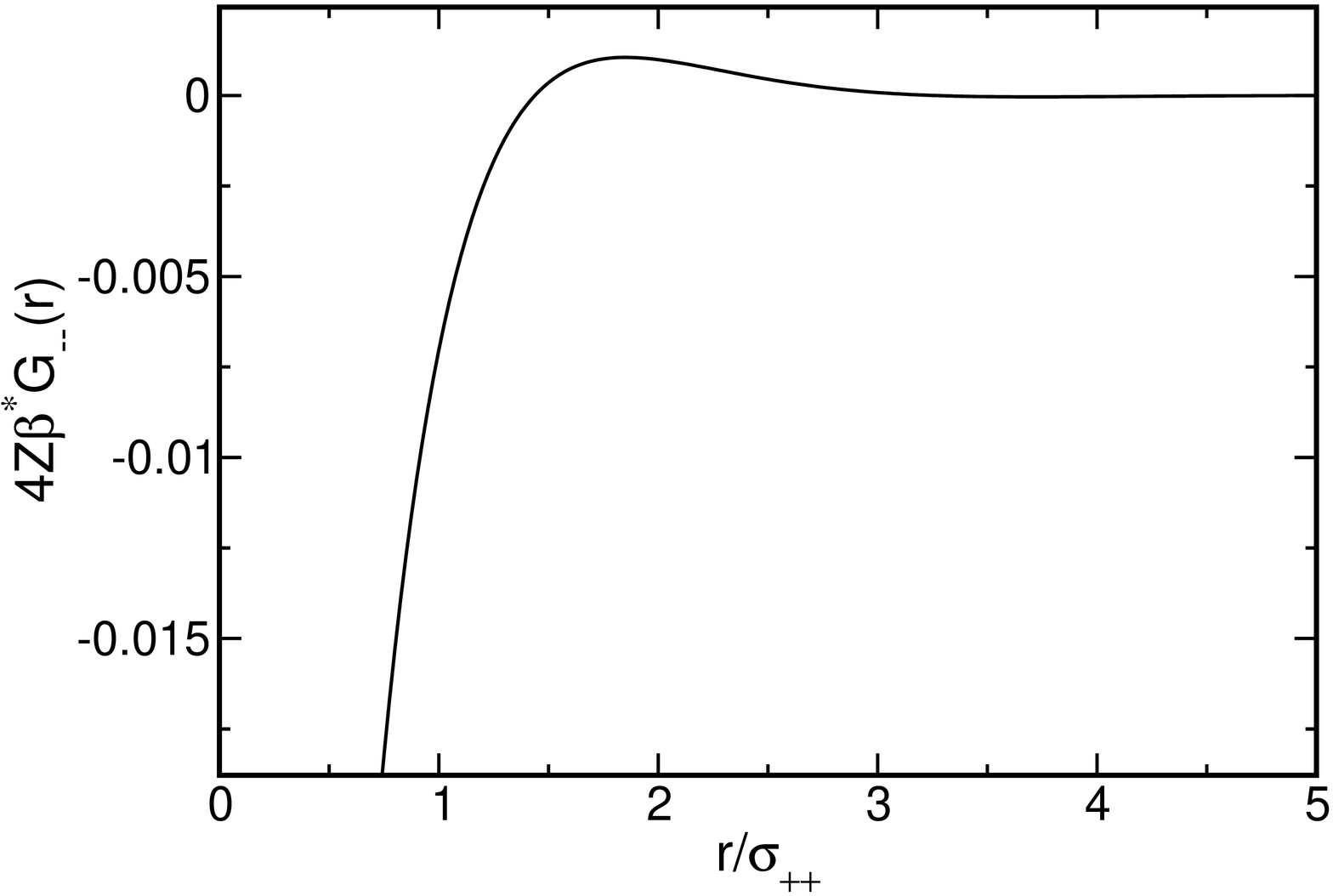}
\caption{Correlation functions $4Z\beta^*G_{\alpha\beta}(r)$ (dimensionless)
for $S=5$, i.e. in the uniform phase at larger densities, but still
far from the transition to the bcc crystal.  The characteristic
lengths in Eq.(\ref{corr++real}) are $\alpha_0=1.45$ and
$\alpha_1=1.72$.  Distance is in units of the particle diameter. As
discussed in sec.~\ref{sec2.2}, results of the mesoscopic theory for
$r\le\sigma$ are not expected to be correct.
}
\end{figure}
\begin{figure}
\includegraphics[scale=0.3,angle=270]{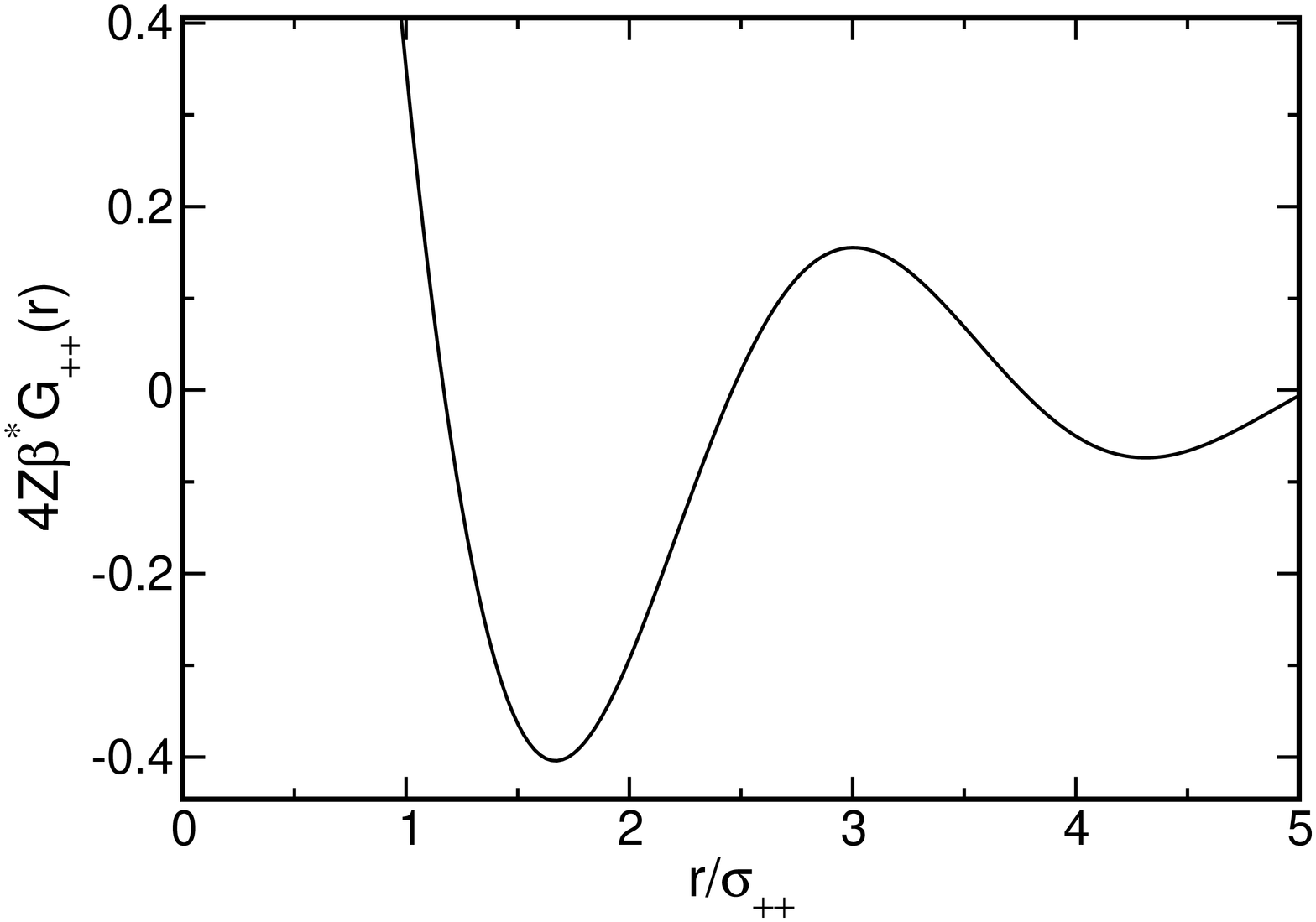}
\includegraphics[scale=0.3,angle=270]{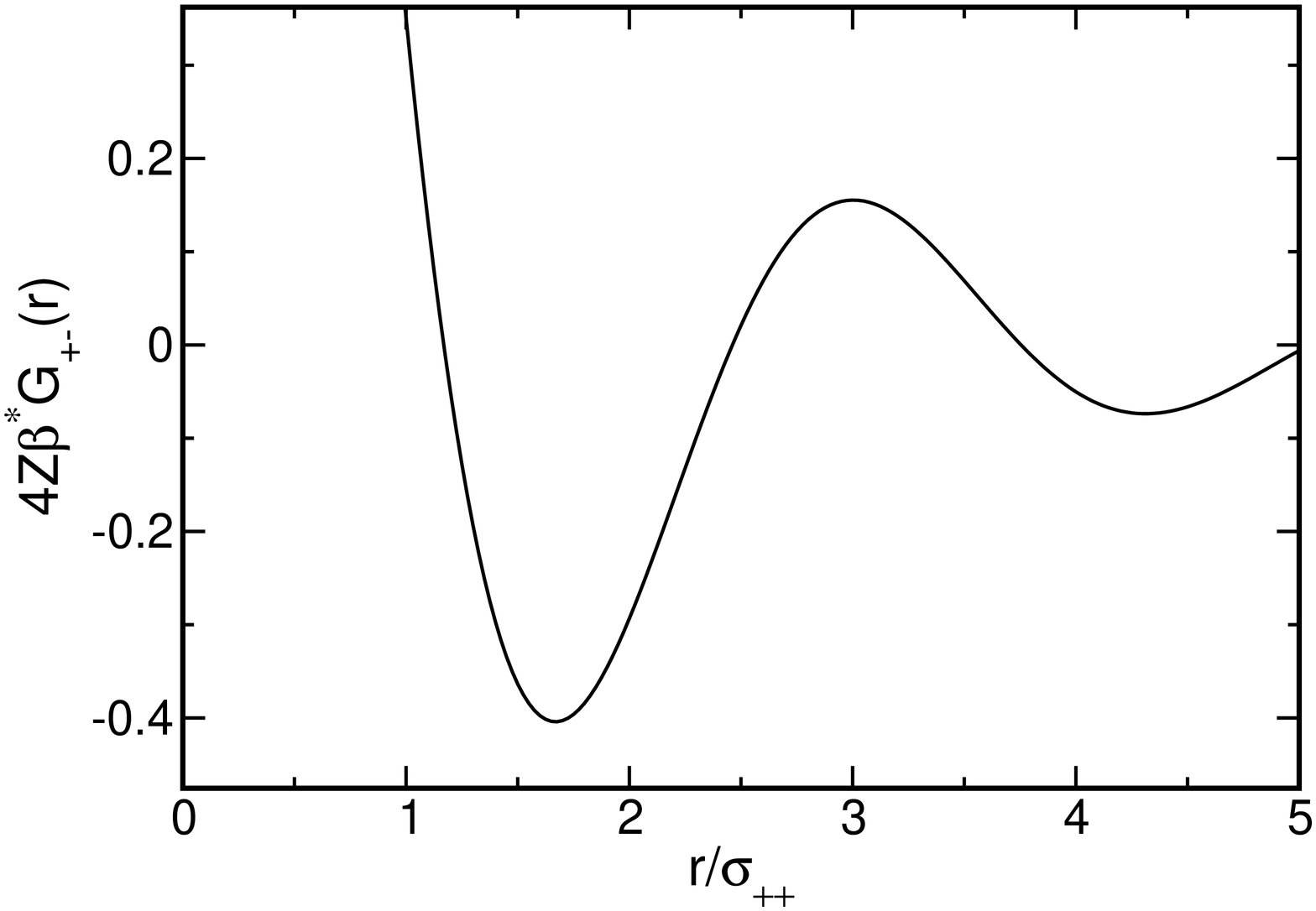}
\includegraphics[scale=0.3,angle=270]{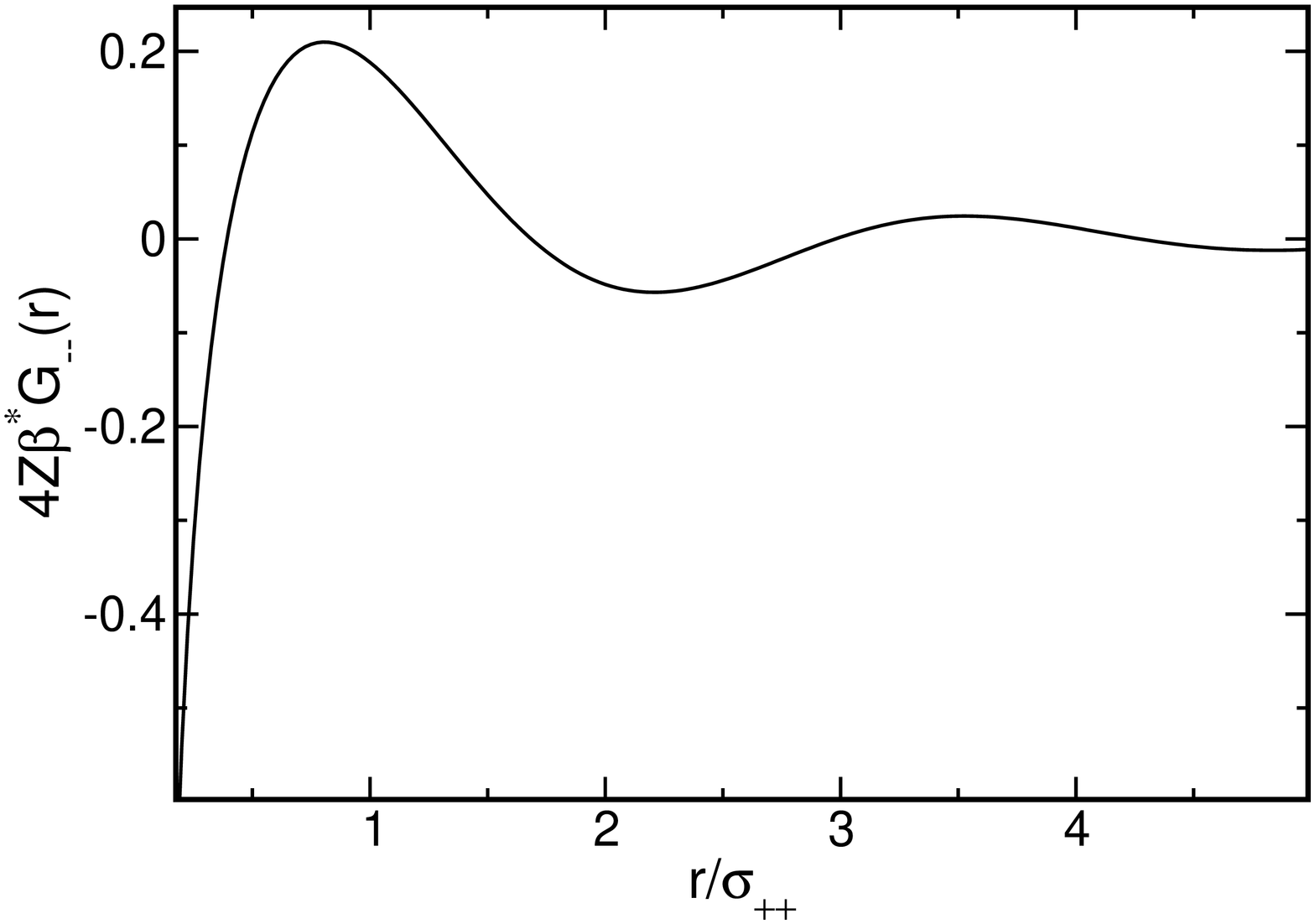}
\caption{Correlation functions $4Z\beta^*G_{\alpha\beta}(r)$ (dimensionless)
for $S=1.712$, i.e. in the uniform phase at the transition to the bcc
crystal. The characteristic lengths in Eq.(\ref{corr++real}) are
$\alpha_0=0.3$ and $\alpha_1=2.43$. Distance is in units of the
particle diameter. As discussed in sec.~\ref{sec2.2}, results of the
mesoscopic theory for $r\le\sigma$ are not expected to be correct.
}
\end{figure}

From Eq.(\ref{monodecay}) it follows that for $r\gg a_2^{-1}$ (where
$a_2>a_1$ and the numbers are well separated) we recover the well
known Yukava-type decay of correlations (see Fig.2), expected for very
dilute systems. In this regime the effective interactions between the
like-charge ions are purely repulsive, as in the DLVO theory.

 In Fig.3 we see a qualitative change in the shape of the correlation
 functions, which in this part of the phase diagram exhibit
 oscillatory decay (\ref{corr++real}). Let us first analyze
 $4Z\beta^*G_{+-}(r)$ and $4Z\beta^*G_{--}(r)$. The
 $4Z\beta^*G_{+-}(r)$ assumes a maximum for $r\approx \sigma_+/2$ and
 then decreases rather slowly for increasing $r$. For
 $2\sigma_+<r<3\sigma_+$ $4Z\beta^*G_{+-}(r)$ is negative, and assumes
 a minimum for $r\approx 2.5\sigma_+$.  At the same time
 $4Z\beta^*G_{--}(r)>0$ for $1.5\sigma_+<r<3\sigma_+$, and assumes a
 positive maximum for $r\approx 1.8\sigma_+$.  This means that the
 counterions are preferably separated by distances
 $1.5\sigma_+<r<3\sigma_+$, i.e. when there is more than enough room
 for a colloid particle to be located between them. The behavior of
 the two correlation functions suggests a tendency for ordering in a
 structure where a diffuse cloud of counterions is formed around the
 colloid particle.  The cloud of counterions extends to the distance
 from the center of the colloid $r\approx 2 \sigma_+$.  For the
 distance from the center of the colloid particle $r>2\sigma_+$ the
 density of the counterions is depleted compared to $\rho_{0-}^*$.
 Let us turn to the $4Z\beta^*G_{++}(r)$. It assumes a positive
 maximum at $r\approx 3\sigma_+$, indicating preferable location of
 the corresponding pair of ions at such distances, consistent with
 formation of the cloud of counterions between them. Note that the
 clouds surrounding the two colloid particles separated by $r\approx
 3\sigma_+$ overlap weakly in a small region around half the distance between
 the particles. The maximum of $4Z\beta^*G_{++}(r)$ in Fig.3 is only
 slightly larger from zero, and the tendency of the colloids to be
 separated by such a distance is very weak.

Let us finally analyze the correlations in the uniform phase at the
coexistence with the bcc crystal. Note first that the phase coexistence 
(\ref{bccT}) 
 occurs quite close to the spinodal line (\ref{spinmacro}),
 where the amplitudes
 of the
 correlation functions diverge, and for $S\to S_b^+$ 
 behave as $\sim(S-S_b)^{-1/2}$ in the Gaussian approximation
 \cite{ciach:03:1} ($S$ is defined in 
Eq.(\ref{S}), and
  $S_b=\pi T^*_b/(24s)$). Thus, the amplitudes of the correlation 
functions are large at the coexistence with the
 crystal, and the fluid phase  is strongly structured. 
 To investigate this structure in more
 detail, consider first 
$4Z\beta^*G_{+-}(r)$ and $4Z\beta^*G_{--}(r)$. We find a
rather large value of $4Z\beta^*G_{+-}(r)$ for
$r\approx\sigma_+/2$, namely $4Z\beta^*G_{+-}(\sigma_{+-})\approx 0.737 $,
and a rather fast decay of  $G_{+-}(r)$
for increasing $r$; $4Z\beta^*G_{+-}(r)<0$ for $1.2\sigma_+<r<2.5\sigma_+$. 
The counterion
correlation function $4Z\beta^*G_{--}(r)$
 assumes a maximum for $r\approx
\sigma_+$, i.e. when the counterions are located at the opposite 
sides of the colloid-particle surface. From the above observations we
can deduce that the cloud of counterions becomes much denser and
thiner, and is closely attached to the particles.  Consider now
$4Z\beta^*G_{++}(r)$.  Positive value of $4Z\beta^*G_{++}(\sigma_+)$
suggests effective attraction between the like charge macroions at the
distance of the closest approach. Similar result was also found in
Ref.\onlinecite{cheong:03:0}. This can only be possible if the point-like
counterions are attached to the colloid surface, consistent with the
formation of a thin and dense layer of counterions around each colloid
particle.  The subsequent, positive maximum of $4Z\beta^*G_{++}(r)$
occurs at $r\approx 3\sigma_+$. This maximum is much higher than away
from the phase transition (Fig.3), and suggests rather strong
tendency for location of colloids at such distances. Note that the clouds of
counterions around the colloid particles separated by the distance
$r\approx 3\sigma_+$ do not overlap.  The second maximum of
$4Z\beta^*G_{+-}(r)$ at $r\approx 3\sigma_+$, i.e. at a similar
distance, shows again that each colloid particle is
surrounded by a dense and thin cloud of counterions.

The
correlation functions $g_{\alpha\beta}$ were obtained in simulations
\cite{cheong:03:0} for $Z=10,\lambda=19$. These asymmetry parameters
are too small for our asymptotic regime $\lambda,Z\to
\infty$, and the correction terms in 
Eqs.(\ref{corgauss})-(\ref{corgauss+-}) may well be of the same order
as the terms which in the asymptotic regime dominate. This may be an
important source of discrepancy between our theory and
simulations. Also, we only obtained the Gaussian correlation functions
in the local-density approximation.  The main discrepancy between our
results and the results of simulations concerns the positions of the
maxima of the correlation functions. Note, however that in experiments
the colloidal crystals are formed in the case of very strong size and
charge asymmetry \cite{arora:98:0}, two-three orders of magnitude
larger than studied in simulations \cite{cheong:03:0}. 
\section{Summary and discussion}
We have developed a mesoscopic theory for the PM with arbitrary size-
and charge asymmetry. Our theory allows for systematic studies of
phase transitions and structure for any charge- and size ratio. The
results obtained in MF and Gaussian approximations can be improved by
adding fluctuation-corrections obtained in perturbation theory.
 
Explicit results for phase transitions and structure
in the uniform phase were obtained in the colloid limit ($\lambda\to
\infty$ first, and next the asymptotic regime of $Z\to \infty$ with
$n_p=O(1/Z)$ is considered) in the MF approximation. We found a
coexistence of a very dilute phase with the bcc crystal formed by the
colloid particles. The lattice constant was found to be $a\approx
3.6\sigma_+$. Very strong dilution of colloids in the 'gas' phase,
structure of the crystalline phase and the lattice constant agree with
experimental results. The correlation functions $G_{\alpha\beta}(r)$
for density deviations of the species $\alpha,\beta=\pm$ at the
distance $r$ show the known monotonic decay for large values of $S$
defined in Eq.(\ref{S}) (high temperatures $T^*$ and/or low densities
$s$). For decreasing $S$ the short-range order in the uniform phase
increases.

 In this work the analysis of the colloid limit is restricted to the
 MF approximation. Inclusion of fluctuations will certainly change the
 quantitative results, in particular the location of the phase
 transition. We expect that the fluctuations do not play a dominant
 role in the colloid limit, but the role of fluctuations certainly
 deserves attention in future works. In the full theory two spinodal
 lines occur, and coupling between the fields $\tilde\rho^*_+$ and
 $\tilde\rho^*_-$ in $\Delta
\Omega^{MF}$ may lead to an increased role of fluctuations.  By
analogy with the RPM \cite{ciach:03:0,ciach:00:0} we expect that for
not too large values of $\lambda,Z $ and/or for volume fractions
larger than $\sim 1/Z$, fluctuations may induce significant shifts of
the spinodal lines, including the change of metastable transitions
into stable ones and vice versa. Hence, beyond MF the crystallization
may be preempted at low concentrations and temperatures by 
the gas-liquid type
separation for certain values of $Z$ and $\lambda$. The role of
fluctuations for different asymmetry parameters will be studied in
future works.

We should emphasize that the foundations of the mesoscopic description
and the asymptotic analysis for large asymmetry are based on
first-principle considerations rather than having been fit to the
results of experiments. Mesoscopic field theories turned out to be
appropriate for a description of a weak ordering, including a weak
crystallization. Because the unit cell of the experimentaly observed
bcc crystal \cite{ise:83:0} is rather large, one may expect that the
corresponding transition is not associated with close packing. The
nearest-neighbor distance in the ordered structure corresponds to the
second maximum in the corresponding correlation function in the
uniform phase close to the phase coexistence. As the results of
 mesoscopic theories are quite accurate down to such distances
\cite{ciach:03:1,leote:94:0,evans:94:0}, it is plausible that in this 
particular case our theory yields correct results on a
semiquantitative level.  Obviously, our mesoscopic field theory has
its limitations, and the structure for distances $\approx \sigma$
cannot be correctly reproduced, as is also the case in the commonly
accepted Landau-Ginzburg-Wilson and Brazovskii theories. Our theory
should be considered as a contribution to the discussion concerning
the thermodynamics and structure in the charged colloidal
systems. Both the experiments and our theory show the formation of the
bcc structure with a large unit cell. To confirm that this is a real
phenomenon it is desirable that microscopic theories and/or
simulations yield similar results.

\begin{acknowledgments}
 The work of AC and WTG was partially funded by the KBN grant No. 1
 P03B 033 26. We also gratefully acknowledge the support
 of the Division of
 Chemical Sciences, Office of the Basic Energy Sciences, Office of
 Energy Research, US Department of Energy.
\end{acknowledgments}

\begin{thebibliography}{10}

\bibitem{stell:76:0}
G. Stell, K. Wu, and B. Larsen, {\it Phys. Rev. Lett.} {\bf 37},  1369  (1976).

\bibitem{fisher:94:0}
M.~E. Fisher, {\it J. Stat. Phys.} {\bf 75},  1  (1994).

\bibitem{outhwaite:04:0}
C.~W. Outhwaite, Cond.-Matter Phys. {\bf 7},  719  (2004).

\bibitem{stell:95:0}
G. Stell, {\it J. Stat. Phys.} {\bf 78},  197  (1995).

\bibitem{onsager:68:0}
L. Onsager, J. Am. Chem.Soc. {\bf 86},  3421  (1968).

\bibitem{stell:68:0}
G. Stell and J. Lebowitz, {\it J. Chem. Phys.} {\bf 49},  3706  (1968), see
  esp. Eqs. (2.9), (2.12) and (2.13).

\bibitem{stell:95:00}
G. Stell,   , sec. 2.2 in Ref. 4.

\bibitem{romero:00:0}
J.~M. Romero-Enrique, G. Orkoulas, A. Panagiotopoulos, and M. Fisher, {\it
  Phys. Rev. Lett.} {\bf 85},  4558  (2000).

\bibitem{yan:01:0}
Q. Yan and J. de~Pablo, {\it Phys. Rev. Lett.} {\bf 86},  2054  (2001).

\bibitem{yan:02:0}
Q. Yan and J. de~Pablo, {\it Phys. Rev. Lett.} {\bf 88},  95504  (2002).

\bibitem{panag:02:0}
A. Panagiotopoulos and M. Fisher, {\it Phys. Rev. Lett.} {\bf 88},  45701
  (2002).

\bibitem{cheong:03:0}
D. Cheong and A. Panagiotopoulos, {\it J. Chem. Phys.} {\bf 119},  8526
  (2003).

\bibitem{rescic:01:0}
J. Rescic and P. Linse, {\it J. Chem. Phys.} {\bf 114},  10131  (2001).

\bibitem{linse:01:0}
P. Linse, {\it Philos. Trans. R. Soc. London, Ser.A} {\bf 359},  853  (2001).

\bibitem{kalyuzhnyi:00:0}
Y.~V. Kalyuzhnyi, M.~F. Holovko, and V. Vlachy, {\it J. Stat. Phys.} {\bf 100},
   243  (2000).

\bibitem{zuckerman:01:0}
D.~M. Zukerman, M.~E. Fisher, and S. Bekiranov, {\it Phys. Rev. E} {\bf 64},
  11206  (2001).

\bibitem{artyomov:03:0}
M.~N. Artyomov, V. Kobelev, and A.~B. Kolomeisky, {\it J. Chem. Phys.} {\bf
  118},  6394  (2003).

\bibitem{aqua:04:0}
J.-N. Aqua and M.~E. Fisher, {\it Phys. Rev. Lett.} {\bf 92},  135702  (2004).

\bibitem{raineri:00:0}
F. Raineri, J. Routh, and G. Stell, {\it J. Phys. IV (France)} {\bf 10},  99
  (2000).

\bibitem{stell:99:0a}
G. Stell,  in {\em New Approaches to Problems in Liquid-State Theory}, edited
  by C. Caccamo, J.-P. Hansen, and G. Stell (Kluwer Academic Publishers,
  Dordrecht, 1999).

\bibitem{stell:92:0}
G. Stell, {\it Phys. Rev. A} {\bf 45},  7628  (1992).

\bibitem{stell:95:0a}
G. Stell,   , sec.2.1.2 in Ref. 4.

\bibitem{ciach:00:0}
A. Ciach and G. Stell, {\it J. Mol. Liq.} {\bf 87},  253  (2000).

\bibitem{ciach:02:0}
A. Ciach and G. Stell, {\it Physica A} {\bf 306},  220  (2002).

\bibitem{ciach:05:0}
A. Ciach and G. Stell, {\it Int.J. Mod. Phys. B} {\bf 21},  3309  (2005).

\bibitem{ciach:04:0}
A. Ciach and G. Stell, {\it Phys. Rev. E} {\bf 70},  16114  (2004).

\bibitem{ciach:04:1}
A. Ciach, {\it Phys. Rev. E} {\bf 70},  046103  (2004).

\bibitem{ise:83:0}
N. Ise {\it et~al.}, {\it J. Chem. Phys.} {\bf 78},  536  (1983).

\bibitem{ise:99:0}
N. Ise, T. Konish, and B. Tata, {\it Langmuir} {\bf 15},  4176  (1999).

\bibitem{arora:98:0}
A.~K. Arora and B.~V.~R. Tata, Adv. Colloid Interface Sci. {\bf 78},  49
  (1998).

\bibitem{blaaderen:05:0}
M. Leunissen {\it et~al.}, Nature {\bf 437},  235  (2005).

\bibitem{tata:97:0}
B.~V.~R. Tata, E. Yamahara, P.~V. Rajamani, and E. Ise, {\it Phys. Rev. Lett.}
  {\bf 78},  2660  (1997).

\bibitem{ito:94:0}
K. Ito, H. Yoshida, and N. Ise, {\it Science} {\bf 263},  66  (1994).

\bibitem{derjaguin:41:0}
B. Derjaguin and L.~D. Landau, {\it Acta Physicochim. URSS} {\bf 14},  633
  (1941).

\bibitem{verwey:43:0}
E.~J.~W. Verwey and J.~T.~G. Overbeek, {\em Theory of the Stability of
  Lyophobic Collids} (Elsevier, Amsterdam, 1948).

\bibitem{warren:00:0}
P.~B. Warren, {\it J. Chem. Phys.} {\bf 112},  4683  (2000).

\bibitem{barbosa:04:0}
M.~C. Barbosa, M. Deserno, C. Holm, and R. Messina, {\it Phys. Rev. E} {\bf
  69},  51401  (2004).

\bibitem{belloni:97:0}
L. Belloni and O. Spalla, {\it J. Chem. Phys.} {\bf 107},  465  (1997).

\bibitem{groot:91:0}
R.~D. Groot, {\it J. Chem. Phys.} {\bf 94},  5083  (1991).

\bibitem{messina:00:0}
R. Messina, C. Holm, and K. Kremer, {\it Phys. Rev. Lett.} {\bf 85},  872
  (2000).

\bibitem{allahyarov:98:0}
E. Allahyarov, I.~D. Amico, and H. L\"owen, {\it Phys. Rev. Lett.} {\bf 81},
  1334  (1998).

\bibitem{lowen:92:0}
H. L\"owen, P.~A. Madden, and J.-P. Hansen, {\it Phys. Rev. Lett.} {\bf 68},
  1081  (1992).

\bibitem{yu:04:0}
Y.-X. Yu, J. Wu, and G.-H. Gao, {\it J. Chem. Phys.} {\bf 120},  7223  (2004).

\bibitem{sogami:84:0}
I.~S. Sogami and N. Ise, {\it J. Chem. Phys.} {\bf 81},  6320  (1984).

\bibitem{roij:99:0}
R. van Roij, M. Dijkstra, and J.-P. Hansen, {\it Phys. Rev. E} {\bf 59},  2010
  (1999).

\bibitem{tamashiro:03:0}
M.~N. Tamashiro and H. Schiessel, {\it J. Chem. Phys.} {\bf 119},  1855
  (2003).

\bibitem{netz:99:0}
R.~R. Netz and H. Orland, Europhys. Lett. {\bf 45},  726  (1999).

\bibitem{netz:00:0}
R.~R. Netz and H. Orland, Eur. Phys.J. E {\bf 1},  67  (2000).

\bibitem{caillol:04:0}
J.~M. Caillol, {\it J. Stat. Phys.} {\bf 115},  1461  (2004).

\bibitem{ciach:01:0}
A. Ciach and G. Stell, {\it J. Chem. Phys.} {\bf 114},  382  (2001).

\bibitem{blum:02:0}
J. Jiang {\it et~al.}, {\it J. Chem. Phys.} {\bf 116},  7977  (2002).

\bibitem{brazovskii:75:0}
S.~A. Brazovskii, {\it Sov. Phys. JETP} {\bf 41},  8  (1975).

\bibitem{evans:94:0}
R. Evans, R.~L. de~Carvalho, J.~R. Henderson, and D.~C. Hoyle, {\it J. Chem.
  Phys.} {\bf 100},  591  (1994).

\bibitem{leote:94:0}
R.~L. de~Carvalho and R. Evans, {\it Mol. Phys.} {\bf 83},  619  (1994).

\bibitem{stell:99:0}
G. Stell,  in {\em New Approaches to Problems in Liquid-State Theory}, edited
  by C. Caccamo, J.-P. Hansen, and G. Stell (Kluwer Academic Publishers,
  Dordrecht, 1999).

\bibitem{ciach:03:0}
A. Ciach and G. Stell, {\it Phys. Rev. Lett.} {\bf 91},  60601  (2003).

\bibitem{shim:05:0}
Y. Shim, M.~Y. Choi, and H.~J. Kim, {\it J. Chem. Phys.} {\bf 122},  044510
  (2005).

\bibitem{panag:99:0}
A.~Z. Panagiotopoulos and S. Kumar, {\it Phys. Rev. Lett.} {\bf 83},  2981
  (1999).

\bibitem{diehl:03:0}
A.Diehl and A.Z.Panagiotopoulos, {\it J. Chem. Phys.} {\bf 118},  4993  (2003).

\bibitem{diehl:05:0}
A. Diehl and A.~Z. Panagiotopolous, {\it Phys. Rev. E} {\bf 71},  046118
  (2005).

\bibitem{bresme:00:0}
C.~V. F.Bresme and J.L.F.Abascal, {\it Phys. Rev. Lett.} {\bf 85},  3217
  (2000).

\bibitem{vega:03:0}
C. Vega, J. Abascal, C. McBride, and F. Bresme, {\it J. Chem. Phys.} {\bf 119},
   964  (2003).

\bibitem{ciach:01:1}
A. Ciach and G. Stell, {\it J. Chem. Phys.} {\bf 114},  3617  (2001).

\bibitem{lebowitz:64:0}
J.~L. Lebowitz and J.~S. Rowlinson, {\it J. Chem. Phys.} {\bf 41},  133
  (1964).

\bibitem{ciach:03:1}
A. Ciach, W.~T. G\'o\'zd\'z, and R.Evans, {\it J. Chem. Phys.} {\bf 118},  3702
   (2003).

\bibitem{leibler:80:0}
L. Leibler, Macromolecules {\bf 13},  1602  (1980).

\bibitem{ciach:05:1}
A. Ciach, W.~T. G\'o\'zd\'z, and G. Stell, to be published  .

\bibitem{blum:75:0}
L. Blum, Mol. Phys. {\bf 30},  1529  (1975).

\bibitem{blum:77:0}
L. Blum and J.~S. Hoye, {\it J. Phys. Chem.} {\bf 81},  1311  (1977).

\bibitem{hiroike:77:0}
K. Hiroike, Mol. Phys. {\bf 33},  1195  (1977).

\bibitem{yukhnovskii:58:0}
I. Yukhnovskii, Sov.Phys. JETP {\bf 34},  263  (1958).

\bibitem{yukhnovskii:87:0}
I.~R. Yukhnovskii, {\em Phase Transitions of the Second Order, Collective
  Variable Methods} (World Scientific, Singapore, 1978).

\bibitem{patsahan:04:0}
O. Patsahan, Cond.Mat.Phys. {\bf 7},  35  (2004).

\bibitem{hoye:97:0}
J. Hoye and G. Stell, {\it J. Stat. Phys.} {\bf 89},  177  (1997).

\end{thebibliography}


\end{document}